\documentclass[3p,times,twocolumn]{elsarticle}
 \biboptions{comma,sort&compress}
 
\usepackage{graphicx}
\usepackage{amsmath}
\usepackage{here}
\usepackage{ecrc}
\volume{00}

\firstpage{1}

\journalname{Nuclear and Particle Physics Proceedings}
\runauth{Stephan Narison}

\jid{nppp}
\jnltitlelogo{Nuclear and Particle Physics Proceedings}
\usepackage{amssymb}
\usepackage[figuresright]{rotating}

\def\beq{\begin{equation}}
\def\eeq{\end{equation}}
\def\bea{\begin{eqnarray}}
\def\eea{\end{eqnarray}}
\def\bq{\begin{quote}}
\def\eq{\end{quote}}

\def\nnb{\nonumber}
\def\ga{\left(}
\def\dr{\right)}

\def\nnb{\nonumber}
\def\la{\langle}
\def\ra{\rangle}
\def\nin{\noindent}
\def\ba{\vspace*{-0.2cm}\begin{array}}
\def\ea{\end{array}\vspace*{-0.2cm}}

\def\d{$\diamond~$}

\def\als{\alpha_s}

\def\g2{ \la\alpha_s G^2 \ra}
\def\g3{g^3f_{abc}\la G^aG^bG^c \ra}
\def\ggg4{\la\als^2G^4\ra}

\begin{document}

\begin{frontmatter}

%%
%%%%%%%%%%%%%%%%%%%%%%%%%%%%%%%%%%%%%%%%%%%%%%%%%
%\begin{document}
\title{
%$\la g^3f_{abc} G^aG^bG^c\ra$
%$\la g^3 f_{abc}G^3\ra$ 
% 
%$\alpha_s $, $\la \alpha_sG^2\ra$, $\overline{m}_{c,b}$ and $f_{B_c}$
%from  relativistic heavy quark sum rules$^*$} 
QCD spectral sum rules 2022\,$^*$} 
 
 \cortext[cor0]{Compact Review talk presented at QCD22, 25th International Conference in QCD (4-7 july 2022,
  Montpellier - FR) and at HEPMAD 22 , 12th HEP International Conference (10-16 october 2022, Antananarivo - MG)}

 \author[label1]{Stephan Narison
 \corref{cor1} 
 }
   \address[label1]{Laboratoire
Univers et Particules , CNRS-IN2P3,  
Case 070, Place Eug\`ene
Bataillon, 34095 - Montpellier Cedex 05, France\\
and\\
Institute of High-Energy Physics of Madagascar (iHEPMAD), 
University of Ankatso,
Antananarivo 101, Madagascar
}
%\cortext[cor1]{ICTP-Trieste  researcher consultant for Madagascar.}
\ead{snarison@yahoo.fr}

\pagestyle{myheadings}
\markright{ }
\begin{abstract}
\noindent
We present a compact review of the status of QCD spectral sum rules until 2022. We  emphasize the recent progresses for determining the QCD input parameters ($\alpha_s$, running quark masses, quark and gluon condensates) where their correlations have been taken into account.  Some selected phenomenological  uses of the sum rules  in different channels (light and heavy quarks, gluonia/glueballs, hybrids and four-quark states) are briefly reviewed and commented. The estimate of the $1^{-+}$ light hybrid mass is revised which confirms the hybrid nature of the $\pi_1(1600)$ but not the $\pi_1(2050)$.   %Some optimal results based on stability criteria are quoted. 

%% keywords
%\ccode{Pac numbers: 11.55.Hx, 12.38.Lg, 13.20-Gd, 14.65.Dw, 14.65.Fy, 14.70.Dj}  
\end{abstract}
\end{frontmatter}
%%%%%%%%%%%%%%%%%%%%%%%%%%%%%%%%%%
%\end{document}
%%%%%%%%%%%%%%%%%%%%%%%%%%%%%%%%%%
%\vspace*{-1.5cm}
\section{Pre-QCD current algebra sum rules}
\vspace*{-0.2cm}
 %\nin
%%%%%%%%%%%%%%%%%%%%%%%%%%%%%%%%%%%
\d Since the famous pre-QCD Weinberg\,\cite{WEIN} (resp. Das, Mathur, Okubo (DMO)\,\cite{DMO}) sum rules of the 1967 for predicting the axial vector $A_1$ mass : $M_{A_1}\simeq\sqrt{2}M_\rho$ assuming the asymptotic realization of chiral $SU(n)_L\otimes SU(n)_R$ (resp. flavour $SU(n)_{L+R}$) symmetry, spectral sum rules have been applied successfully to study
the masses and decay constants of hadrons.

\d The breaking of the previous sum rules by the quark masses and QCD coupling $\alpha_s$ have been studied in Ref.\,\cite{FLORATOS,DMOQCD} where it has been shown that the 1st Weinberg sum rule is only broken by the running light quark mass to order $\alpha_s$ while the 2nd  Weinberg and DMO sum rules are broken to leading order (LO). 

%%%%%%%%%%%%%%%%%%%%%%%%%%
\vspace*{-0.3cm}
\section{The SVZ sum rules}
\begin{figure}[hbt]
\begin{center}
\includegraphics[width=7.5cm]{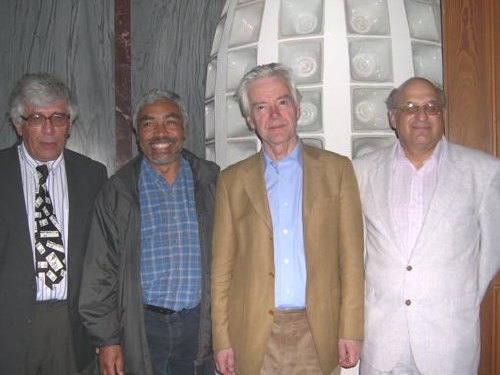}
\vspace*{-0.15cm}
\caption{\footnotesize  Left to right : Arakady Vainshtein, SN, Valya Zakharov, Mikhael Shifman at the Munich conference (2006). }
\label{fig:svz}
\end{center}
\vspace*{-0.5cm}
\end{figure}

\vspace*{-0.20cm}
%%%%%%%%%%%%%%%%%%%%%%%%%%
\d Besides the Shifman-Vainshtein-Zakharov (SVZ) (see Fig.\,\ref{fig:svz}) theoretical improvement of the perturbative QCD expression introduced in 1979\,\cite{SVZa,ZAKA}\,\footnote{For reviews, see e.g. the books\,\cite{SNB1,SNB2} and references therein and the recent ones in Refs.\cite{SNREV1,SNbc15}.} which we shall discuss in the next section,  the phenomenological success of QCD (spectral) sum rules or SVZ sum rules  comes from the improvemnt of the usual dispersion relation which is the bridge between the high-energy QCD expression and the measurable spectral function at low energy  from e.g.  the relation between the electomagnetic spectral function and the $e^+e^-\to hadrons$ cross-section via the optical theorem. 
%%%%%%%%%%%%%%%%%%%%%%%%%%%%%%%%%%%%%%%
\vspace*{-0.25cm}
%%%%%%%%%%%%%%%%%%%%%%%%%%%%%%%%%%%%%%%%%

\d This improvement has been been achieved by working with large number $n$ of derivatives and large value of the $Q^2$ momentum transfer $Q^2$ but taking their ratio $\tau\equiv n/Q^2$ finite leading to the so-called Borel/Laplace or Exponential sum rules (LSR) and their ratios,\footnote{Non-relativistic vesion of this sum  rule has been discussed by\,\cite{BELLa,BERTa}, while the inclusionof the PT $\alpha_s$ correction to the QCD expression has shown that it has the property of an inverse Laplace transform\,\cite{SNR} though the name LSR.} :
\bea
{\cal L}_0^c(\tau,\mu)&\equiv&\lim_ {\begin{tabular}{c}
$Q^2,n\to\infty$ \\ $n/Q^2\equiv\tau$
\end{tabular}}
\frac{(-Q^2)^n}{(n-1)!}\frac{\partial^n \Pi}{ ( \partial Q^2)^n}\nnb\\
&=&\int_{t>}^{t_c}dt~e^{-t\tau}\frac{1}{\pi} \mbox{Im}\Pi_H(t,\mu)~,
\nnb\\
 {\cal R}^c_{10}(\tau)&\equiv&\frac{{\cal L}^c_{1}} {{\cal L}^c_0}= \frac{\int_{t>}^{t_c}dt~e^{-t\tau}t\, \mbox{Im}\Pi_H(t,\mu)}{\int_{t>}^{t_c}dt~e^{-t\tau} \mbox{Im}\Pi_H(t,\mu)},~~~~
\label{eq:lsr}
\eea
where $\tau$ is the LSR variable, $t>$   is the hadronic threshold.  Here $t_c$ is  the threshold of the ``QCD continuum" which parametrizes, from the discontinuity of the Feynman diagrams, the spectral function  ${\rm Im}\,\Pi_H(t,m_Q^2,\mu^2)$.  

\d For heavy quark systems, another successful sum rules 
are the $Q_0^2$-moments sum rules (MSR) and their ratios\,:
\bea
 {\cal M}^c_n(Q_0^2,\mu)&=&\frac{1}{n!}\ga\frac{\partial}{\partial q^2}\dr^n\Pi_H(q^2,m_Q^2)\vert_{q^2=-Q_0^2}\nnb\\
 &=&\int_{t>}^{t_c}\hspace*{-0.cm}\frac{dt}{(t+Q^2_0)^n}\frac{1}{\pi} \mbox{Im}~\Pi_H(t,\mu)~,\nnb\\
 {r}^c_{n\,n+p}&=&\frac{{\cal M}^c_{n}} {{\cal M}^c_{n+p}}~~: p=1,2,\dots,
 \eea
 where the $m_Q$ heavy quark mass has been exploited for doing the OPE in terms of $1/m_Q$ expansion. $Q_0^2=0, m_Q^2,..$ is a free chosen scale and $n$ is the degree of moments. 
 
\d  $\Pi_H(t,m_Q^2,\mu^2)$ is the generic two-point correlator defined as :
 \beq
\hspace*{-0.6cm} \Pi_H(q^2)=i\hspace*{-0.1cm}\int \hspace*{-0.15cm}d^4x ~e^{-iqx}\la 0\vert {\cal T} {\cal O}_H(x)\ga {\cal O}_H(0)\dr^\dagger \vert 0\ra~.
 \label{eq:2-point}
 \eeq
 
\d $ {\cal O}_H(x)$ can be the interpolating local currents:

-- Quark bilinear local current $\bar \psi_1\Gamma_{12}\psi_2$ for mesons. 
$\Gamma_{12}$ is any Dirac matrices which specify the quantum numbers of the corresponding meson state (and its radial excitations),

-- Quark trilinear local current $\psi_1\psi_2\psi_3$ for baryons,

--   Four-quark $(\bar \psi_1\Gamma_{12}\psi_2)(\bar \psi_3\Gamma_{34}\psi_4)$ or diquark anti-diquark $(\bar \psi_1\Gamma_{12}\bar\psi_2)(\psi_3\Gamma_{34}\psi_4)$ local current for molecules or tetraquark states. 

-- Gluon local currents $G^2, G^3\cdots$ for gluonia/glueball states,

-- Quark-gluon local currents $\bar\psi G\psi, \bar\psi G^2\psi\cdots $ for hybrid states.

\d The previous sum rule improvements enhance the low energy contribution to the spectral integral which is accessible experimentally. In the often case where the data
on the spectral function are not available, one usually parametrizes it via the minimal duality ansatz :
\vspace*{-0.2cm}
\bea
\hspace*{-0.25cm}
\frac{1}{\pi}\mbox{ Im}\Pi_H(t)&\simeq&  f^2_HM_H^{2d}\delta(t-M^2_H)\nnb\\
  &+&
  ``\mbox{QCD continuum}" \theta (t-t_c),
\label{eq:duality}
\eea
 in order to predict the masses and couplings of the lowest ground state and in some case the ones of its radial excitations.  $d$ depends on the dimension of the current, $f_H$ is the hadron decay constant  normalized as $f_\pi=$ 132 MeV. Its accuracy has been tested in various light and heavy quark channels $e^+e^-\to \rho,J/\psi, \Upsilon,\dots$ where complete data are available\,\cite{SNB1,SNB2} and in the $\pi$-pseudoscalar channel where an improved parametrization of the $3\pi$ channel within chiral perturbation theory has been used\,\cite{BIJNENS}. Within a such parametrization, the ratio of sum rules is used to extract the mass of the lowest ground state as it is equal to its square. However, this analysis cannot be done blindly without studying / checking the  moments which can violate positivity for some values of the sum rule variables ($\tau, n, Q_0^2$)  though their ratio can lead to a positive number identified with the hadron mass squared.  

\d Within the duality ansatz paametrization of the spectral function, the ratios ${\cal R}_n^c(\tau)$ and  $r^c_{n,n+1}$ is approximately equal to the hadron mass squared while the double ratio of sum rule (DRSR) \,\cite{DRSR} :
\beq
r_{H'/H}\equiv \frac{{\cal R}_{10}^c(\tau')\vert_{H'}}{{\cal R}_{10}^c(\tau)\vert_{H}}\simeq \frac{M_{H'}^2}{M_{H}^2}
\eeq
can be used to get the meson $H$ and $H'$ mass-splitting, like e.g. the one due to $SU(3)$ breakings, provided that the optimal value: $\tau'_0\simeq \tau_0$. A similar quantity can be used for the heavy quark moments.
%%%%%%%%%%%%%%%%%%%%%%%%%%%%%%%%%%%%
 \vspace*{-0.3cm}
\section{Some other type of sum rules}
\vspace*{-0.15cm}
%%%%%%%%%%%%%%%%%%%%%%%%%%%%%%%%%%%%%
\d Alternative to these SVZ sum rules is the local Finite Energy Sum Rule (FESR) :
%%%%%%%%%%%%%%%%%%%%%%%%%%%%%%%%%%%
\beq
 {\cal F}^c_n(\mu)=\int_{t>}^{t_c}\hspace*{-0.cm}dt\,t^n\frac{1}{\pi} \mbox{Im}~\Pi_H(t,\mu)~,
\eeq
where $n\geq 0$ is the degree of the sum rule. Contrary to LSR, FESR emphasizes the role of higher masses radial excitations to the integral. This sum rule can also be derived from LSR by using a $\tau\to 0$ expansion. FESR can be used to fix the value of the QCD continuum threshold $t_c$ from its dual constraint with the mass and decay constant of the ground state\,\cite{FESR1,FESR2}. However, the accuracy of the result is destroyed when  $n$ increases due to the needs of more information for the spectral function near the cut on the real axis. Therefore, the FESR should be used with a great care for high $n$ values. 
%%%%%%%%%%%%%%%%%%%%%%%%%%%%%%%%%%%
\d Another sum rule is the Gaussian sum rule\,\cite{FESR1,FESR2}:
%%%%%%%%%%%%%%%%%%%%%%%%%%%%%%%%%%%
\beq
{G}^c_n(s,\sigma,\mu)=\frac{1}{\sqrt{4\pi\sigma}}\int_{t>}^{t_c}dt~e^{-\frac{(t+s)^2}{4\pi}}\frac{1}{\pi} \mbox{Im}\Pi_H(t,\mu)~,
\eeq
for a Gaussian centered at $s$ with a finite width resolution $\sqrt{4\pi\sigma}$. It has been aslo shown in Ref.\,\cite{FESR1,FESR2} that the LSR can be derived from the Gaussian sum rule using the $\zeta$-prescription.
%%%%%%%%%%%%%%%%%%%%%%%%%%%%%%%%%%
\d Another sum rule is the $\tau$-like-decay sum rule\,\cite{BNP}: 
%%%%%%%%%%%%%%%%%%%%%%%%%%%%%%%%%%%
\bea
&&R_\tau(M_\tau)= \int_{0}^{M_\tau^2}\frac{ds}{M_\tau^2} \ga 1-\frac{s}{M_\tau^2}\dr^2\times\nnb\\
&&\Bigg{\{}\ga 1+\frac{2s}{M_\tau^2}\dr {\rm Im}\Pi_H^{(1)}(s) + {\rm Im}\Pi_H^{(0)}(s)\Bigg{\}},
\eea
for a spin one and zero hadronic final states or its moment\,\cite{LEDI}. Contrary to the usual FESR, it has the advantage to have the threshold suppression factor near the real cut. This property has enabled to extract with high level of accuracy the QCD coupling at the $M_\tau$ scale.
 %%%%%%%%%%%%%%%%%%%%%%%%%%%%%%%%%%%%%%%%%%%
 \vspace*{-0.25cm}
\section{The SVZ - OPE\label{sec:ope}}
\vspace*{-0.25cm}
%%%%%%%%%%%%%%%%%%%%%%%%%%%%%%%%%%%%%%%%%%%
\d According to SVZ, the RHS of the two-point function can be evaluated in QCD within the Operator Product Expansion (OPE) provided that $ \Lambda^2\ll Q^2\equiv -q^2, m_Q^2$. In this way, it reads\,:
\beq
\Pi_H(q^2,m_Q^2,\mu)=\sum_{D=0,2,..}\hspace*{-0.25cm}C_{D}(q^2,m_Q^2,\mu)\la O_{D}(\mu)\ra~, 
\eeq
where, in addition to the usual perturbative QCD contribution (unit operator), one can add the ones due to non-perturbative gauge invariant quark and gluon condensates $\la O_{D}(\mu)\ra$ having a dimension $D$ which have been assumed to  parametrize approximately the not yet under good control QCD confinement.  $C_{D}$ are separated calculable Wilson coefficients in PT-QCD:

\d The usual perturbative (PT) contribution corresponds to $D=0$ while the quadratic quark mass corrections enter via $D=2$. 

\d The quark and gluon condensates entering into the OPE up to $D=6$ are successively the :

-- $D=4$ quark and gluon $ m\la\bar\psi\psi\ra$ and $ \la \alpha_s G^2\ra$, 

-- $D=5$ mixed quark-gluon: $\la\bar\psi\sigma^{\mu\nu}\frac{\lambda_a}{2}G_{\mu\nu}^a\psi\ra$

-- $D=6$ four-quark and three-gluon:  $\la\bar\psi  \Gamma_1\psi \bar\psi \Gamma_2\psi\ra$, \hspace*{1cm} $ \la g^3f_{abc} G^a_{\mu\rho}G^{b,\rho}_{\nu}G^{c,\nu}_\rho\ra$. 

\d The $D=4$ condensates
$m\la\bar\psi\psi\ra$ and the part of the trace of the energy-momentum transfer\,: $\theta^\mu_\mu\vert_g\equiv m\gamma \la\bar\psi\psi\ra +(1/4)\beta \la  G^a_{\mu\nu}G_a^{\mu\nu}\ra$ are known to be subtraction $\mu$-independent where $\gamma,\, \beta$ are the quark mass anomalous dimension and Callan-Symanzik $\beta$-function.  

-- The $\la\bar\psi\psi\ra$ condensate can be deduced from the well-known Gell-Mann, Oakes, Renner relation\,\cite{GMOR}:
\vspace*{-0.15cm}
\beq
(m_u+m_d)\la\bar\psi_u\psi_u+\bar\psi_d\psi_d\ra  = -m_\pi^2f_\pi^2,
\eeq
\vspace*{-0.15cm}
once the running light quark mass is known ($m_\pi, f_\pi =132$ MeV are the pion mass and decay constant) or directly from light baryon sum rules\,\cite{DOSCHSN}. With the value of the running $u,d$ quark masses given in Table\,\ref{tab:param}, one can deduce the value of  $\la\bar\psi\psi\ra$ in this table. 

-- The original value of the $D=4$ gluon condensate $\la \alpha_s G^2\ra=0.04$ GeV$^4$ of SVZ\,\cite{SVZa,RRY} has been claimed\,\cite{BELLa,BERTa} from charmonium sum rules and  Finite Energy Sum Rule (FESR) for $e^+e^-\to I=1$\, hadrons\,\cite{FESR1,FESR2} to be underestimated. Recent analysis from $e^+e^-\to hadrons$, $\tau$-decay and charmonium confirm these claims
(see the determinations in Table 1 of\,\cite{SNparam}). The present updated  average is (see Table\,\ref{tab:param}):
\vspace*{-0.15cm}
\beq
\la \alpha_s G^2\ra=(6.49\pm 0.36)\times 10^{-2}~{\rm GeV}^4. 
\eeq
\vspace*{-0.15cm}
The renormalization of higher dimension  condensates have been studied in\,\cite{SNTARRACH} where:

\d The $D=5$ quark-gluon mixed condensate usually parametrized as $g\la\bar \psi G\psi\ra=M_0^2\la \bar \psi\psi\ra$  mixes under renormalization and runs as $(\alpha_s)^{1/(6\beta_1)}$ in the chiral limit $m=0$. The scale :
\vspace*{-0.15cm}
\beq
M_0^2=0.8(2)~{\rm GeV}^2,
\eeq
\vspace*{-0.15cm}
has been phenomenologically estimated from light baryons\,\cite{IOFFE,DOSCH,PIVOm} and heavy-light mesons\,\cite{SNhl} sum rules.

\d The $D=6$ four-quark condensate mixes under renormalization with some other ones which is not compatible with the vacuum saturation assumption used by SVZ.  Its phenomenological estimate from $\tau$-decays\,\cite{SNTAU}, $e^+e^-\to$ hadrons data\,\cite{LNT}, $\tau$-decay\,\cite{SOLA}, FESR \,\cite{FESR1,FESR2}  and baryon\,\cite{DOSCH} sum rules, leads to:
\vspace*{-0.15cm}
\beq
\rho \alpha_s\la\bar\psi\psi\ra^2\simeq 5.8(9)10^{-4}\,{\rm GeV}^6~:~~~\rho\simeq 2\sim 4~,
\eeq
\vspace*{-0.15cm}
 where $\rho$, indicates the deviation from factorization. 
 
 \d Fixing the ratio $\la \alpha_s G^2\ra/\rho \alpha_s\la\bar\psi\psi\ra^2= 106(12)$ GeV$^{-2}$ as quoted in Ref.\,\cite{SND21} which reduces the analysis to a one-parameter fit, one deduces from LSR\,\cite{SNTAU}:
 \vspace*{-0.45cm}
 \beq
 \la \alpha_s G^2\ra=  (6.1\pm 0.7)10^{-2}~{\rm GeV}^4,
 \eeq
 \vspace*{-0.15cm}
which shows the self-consistency of the previous numbers.  Some other consistency tests can be found in\,\cite{SNTAU}.

\d  The $D=6$ $g^3f_{abc}\la G^aG^bG^c \ra$ condensate does not mix under renormalization and behaves as $(\alpha_s)^{23/(6\,\beta_1)}$, where $\beta_1=-(1/2)(11-2n_f/3)$ is the first coefficient of the $\beta$-function and $n_f$ is number of quark flavours. 
The first improvement of the estimate of the $g^3f_{abc}\la G^aG^bG^c \ra$ condensate was the recent direct determination of 
the ratio of the dimension-six gluon condensate $\la g^3f_{abc} G^3\ra$ over the dimension-four one $\la\alpha_s G^2\ra$ using heavy quark sum rules with the value\,\cite{SNcb1}:
\vspace*{-0.15cm}
\beq
\rho_G\equiv \la g^3f_{abc} G^3\ra/ \la \alpha_s G^2\ra=(8.2\pm 1.0)~{\rm GeV}^2,
\label{eq:rcond}
\eeq
\vspace*{-0.15cm}
which differs significantly from the instanton liquid model estimate\,\cite{NIKOL2,SHURYAK,IOFFE2}. This result may question the validity of a such result. 
Earlier lattice results in pureYang-Mills found:  $\rho_G\approx 1.2$ GeV$^2$\,\cite{GIACO} such that it is important to have new lattice results for  this quantity. Note however, that the value given in Eq.\,\ref{eq:rcond} might also be an effective value of all unknown high-dimension condensates not taken into account in the analysis of \,\cite{SNcb1} when requiring the fit of the data by the truncated OPE if, at that order, the OPE does not converge. We shall see here and in some examples that the effect of  $ \la g^3f_{abc} G^3\ra$ is a small correction at the stability  region where the optimal results are extracted.  
 
\d Usually, the truncation of the OPE up to $D=6$ provides enough information for extracting the masses and couplings of the ground state hadrons with a good accuracy. In many papers, some classes of higher dimension terms up to D=12 !  are added in the OPE. However, it is not clear if such  term gives the dominant contributions compared to the omitted ones having the same dimension.  The size of these high-dimension condensates is not also under a good control due to the eventual violation of the vacuum saturation used for their estimate. 

\d Applied to the previous Weinberg sum rules\,\cite{WEIN}, it has been found\,\cite{FLORATOS} that, in the chiral limit $m_q=0$, the 1st Weinberg sum rule is spontaneously broken by the four-quark condensate to order $\alpha_s\la\bar qq\ra^2$ while for non-zero quark masses, there are contributions from $m_im_j$  and $ m_i\la \bar \psi_i\psi_i\ra$ quark condensate to these Weinberg and DMO sum rules\,\cite{FLORATOS,DMOQCD}. 
%%%%%%%%%%%%%%%%%%%%%%%%%%%%
\vspace*{-0.3cm}
\section{Beyond the SVZ-OPE}
\vspace*{-0.2cm}
%%%%%%%%%%%%%%%%%%%%%%%%%%%%
\subsection*{\d The $D=2$ tachyonic gluon mass}
%%%%%%%%%%%%%%%%%%%%%%%%%%%
\vspace*{-0.1cm}
 The asymptotic behaviour of the PT series is often expected to have an exponential behaviour (Borel sum) according to the large $\beta$-approximation and then alternate signs are expected to be seen at large orders of PT. However, the known calculated terms up to order $\alpha_s^5$ of the vector correlator $D$-function do not yet show such properties. In Refs.\,\cite{CNZa,CNZb,ZAKa,ZAKb}, a phenomenological parametrization of these higher order terms  due to UV-renormalons have been proposed which is parametrized by a tachyonic gluon mass squared contribution.  Its phenomenological value from $e^+e^-\to$ hadrons data\,\cite{SND21}, $\pi$-Laplace sum rule\,\cite{CNZa} and from an analysis of the lattice data of the pseudoscalar $\oplus$ scalar two-point correlators\,\cite{CNZb}\, lead to the average\,\cite{SZ}:
 \vspace*{-0.15cm}
\beq
(\alpha_s/\pi)\lambda^2\simeq -(7\pm 3)\times 10^{-2}\,{\rm GeV}^2.
\eeq
\vspace*{-0.15cm}
 The existence of this $D=2$ term not present in the standard OPE (absence of gauge invariant $D=2$ term) has raised some vigourous (unjustifed and emotional) reactions though its contribution is tiny in the sum rule and $\tau$-decays\,\cite{SNTAU} analyses but it has solved some paradoxical sum rule scale puzzles\,\cite{CNZb}. This $D=2$ term also manifests as a linear term of the heavy quark potentials\,\cite{ADS1} and in the SVZ-expansion\,\cite{ADS3} in some AdS/QCD  models. However, this term is not of InfraRed origin like some other non-perturbative condensates but it is dual to the sum of higher order Ultra-Violet terms of the PT series as shown in\,\cite{SZ} : {\it better the series is known , lesser is the strength of this term which can vanish after some high order
terms of the PT series}.   A such term is dual to a geometric sum of the coefficients of the PT series and its size is consistent with the values of the known coefficients. 

%%%%%%%%%%%%%%%%%%%%
\vspace*{-0.25cm}
\subsection*{\d Small size instantons}
\vspace*{-0.1cm}
%%%%%%%%%%%%%%%%%%%%
Direct instantons are expected to be present in QCD for explaining the $\eta'
-\pi$ mass shift (the so-called U(1)A axial problem\,\cite{WITTEN}). At large $Q^2$, it will be highly suppressed as it can be parametrized by an operator of high dimension $D = 9$\,\cite{SNB1,SNB2}. Its quantitative effect has been discussed in previous QSSR literature and has lead to some controversy\,\cite{SNB1,SNB2}.  
Applied to the $e^+e^-\to I=1~hadrons$ data,  this effect is found to be negligible\,\cite{SND21},
while in the pseudoscalar channel, it gives a prediction  for $\overline m_{ud}(2)\equiv (1/2)(\overline m_u+\overline m_d)(2)=(2.42 \pm 0.16)$ MeV and $\overline m_s(2)=(63.1\pm 3.4)$ MeV lower than the estimate from the standard SVZ OPE of $(3.95\pm 0.28)$ and $(98.5\pm 5.5)$ MeV\,\cite{SNLIGHT}. This instanton effect is not  favoured by the lattice average\,\cite{LATTLIGHT}: $\overline m_{ud}(2)=(3.6\pm0.2)$ MeV and $\overline m_s(2)=(93.8\pm 2.4)$ MeV. However, one also oberves in Ref.\,\cite{SNLIGHT} that the  ratios of masses are (almost) unaffected by the presence of instanton. 
%%%%%%%%%%%%%%%%%%%%
\vspace*{-0.25cm}
\subsection*{\d Duality violation}
\vspace*{-0.1cm}
%%%%%%%%%%%%%%%%%%%
A model of a duality violation for the spectral function has been proposed in Ref.\,\cite{DV} in order to parametrize the oscillations observed in the spectral function controlling $\tau$-decay and $e^+e^-\to I=1~hadrons$  from the data. Typically, it behaves as :
\vspace*{-0.15cm}
\beq
\Delta {\rm Im}\Pi_H^{DV}\sim  \kappa e^{-\gamma t}\sin(\alpha+\beta t)\theta(t-t_c),
\eeq
\vspace*{-0.15cm}
where $\kappa,\gamma, \alpha,\beta$ are fitted parameters not based from first principles which are channel and $t_c$ dependent. 
%In the case of vector currents of light quarks, these parameters are.\,\cite{DV} :
%\bea
%\kappa_V&\simeq& 0.02,~~~~~~\gamma_V\simeq (0.15\sim 0.4),\nnb\\
%\alpha_V&\simeq &1.8,~~~~~~~~\beta_V\simeq 2.1.
%\eea
Within this model, where the contribution is double exponential suppressed in the Laplace sum rule analysis, we expect that in the stability region where the QCD continuum contribution to the sum rule is minimal and where the optimal results in this paper will be extracted, such duality violations is irrelevant in the sum rule analysis.  A similar conclusion has been recently reached in the determination of $\alpha_s$ from $\tau$-decay using a slightly different parametrization of the spectral function\,\cite{TONI}.

%%%%%%%%%%%%%%%%%%%%%%%%%%%%%%%%%%%
\vspace*{-0.5cm}
\section {Optimization procedure for the SVZ sum rule}
\vspace*{-0.15cm}
%%%%%%%%%%%%%%%%%%%%%%%%%%%%%%%%%%
\subsection*{\d The SVZ sum rule window}
\vspace*{-0.1cm}
The second important step in the sum rule analysis is the extraction of the optimal result from the sum rule as, in principle, the Laplace sum rule (LSR) variable $\tau$, the degree of moments $n$ and the QCD contiuum threshold $t_c$ are free external variables. In their original work,  SVZ have ajusted the values of $M^2\equiv 1/\tau$ and $t_c$ using some guessed $\%$ contributions, for finding the sum rule window where the QCD continuum contribution is less than some input number while the ground state one is bigger than some input number and where,  in this sum rule window, the OPE is expected to converge.  The arbitrariness values of these numbers have created some doubts for non-experts on the results from the sum rules, in addition to the ones on the eventual non-correctness of the non-trivial QCD-OPE expressions. Unfortunately, many sum rules practioners continue to use this SVZ criterion. 
%%%%%%%%%%%%%%%%%%%%%%%%%%%%%%%%%%%%%%%%%%
\vspace*{-0.25cm}
\subsection*{\d $\tau$-stability from quantum mechanics and $J/\Psi$}
\vspace*{-0.25cm}
%%%%%%%%%%%%%%%%%%%%%%%%%%%%%%%%%%%%%%%%%%

%%%%%%%%%%%%%%%%%%%%%%%%%%%%%%%%%%%%%%%
\begin{figure}[hbt]
\begin{center}
\includegraphics[width=6.cm]{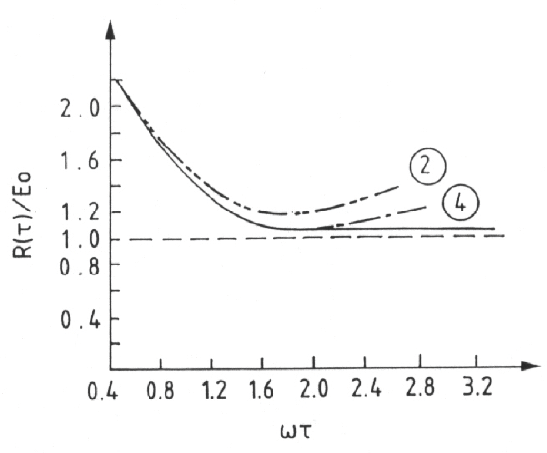}
%\vspace*{-0.7cm}
\caption{\footnotesize  Behaviour of the ground state mass versus the time variable $\tau$ for different truncation of the series from Ref.\,\cite{BELLa,BERTa}. The horizontal line is the exact solution. $\omega$ is the harmonic oscillator frequency.}
\label{fig:oscillo1}
\end{center}
\vspace*{-0.5cm}
\end{figure} 
%%%%%%%%%%%%%%%%%%%%%%%%%%%%%%%%%%%%%%%%%

%%%%%%%%%%%%%%%%%%%%%%%%%%%%%%%%%%%%%%%
\begin{figure}[hbt]
\begin{center}
\includegraphics[width=6.cm]{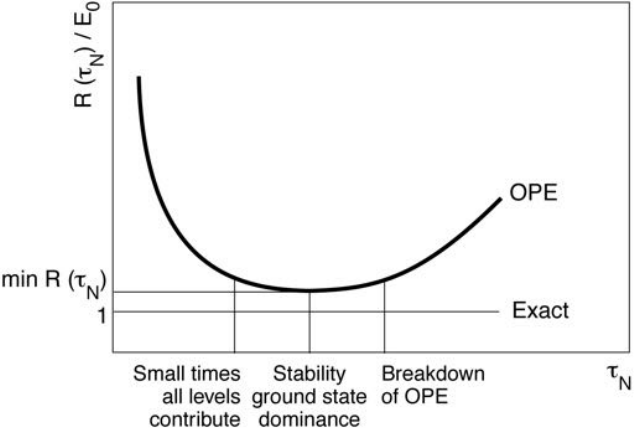}
%\vspace*{-0.7cm}
\caption{\footnotesize  Schematic behaviour of the $J/\psi$ mass  versus the sum rule variable $\tau_N$ from Ref.\,\cite{BELLa,BERTa}. The horizontal line is the experimental mass.} \label{fig:oscillo2}
\end{center}
\vspace*{-0.5cm}
\end{figure} 
%%%%%%%%%%%%%%%%%%%%%%%%%%%%%%%%%%%%%%%%%
 %%%%%%%%%%%%%%%%%%%%%%%%%%%%%%%
%   \vspace*{-1.3cm}
\begin{figure}[hbt]
\begin{center}
\includegraphics[width=7.cm]{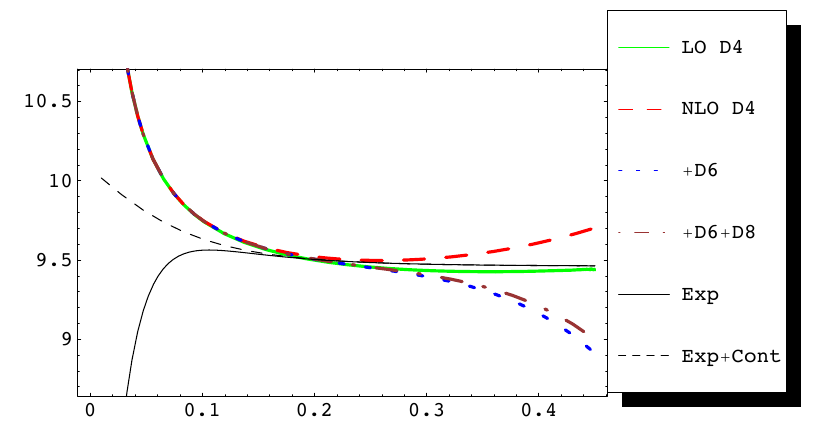}
%\vspace*{-0.7cm}
\caption{\footnotesize  Behaviour of the ratio of moments $\sqrt{{\cal R}^b_0}(\overline{m}_b^2)$ in GeV versus $\tau$ in GeV$^{-2}$ and for $\overline{m}_b(\overline{m}_b) = 4212$ MeV from\,\cite{SNH12}. The black continuous (rep. short dashed) curves are the experimental contribution including (resp. without) the QCD continuum (it is about the $M_{\Upsilon}$). The green (thick continuous) is the PT contribution including the $D=4$ condensate to LO. The long dashed (red) curve is the contribution including the $\alpha_s$ correction to the $D=4$ contribution. The short dashed (blue) curve is the QCD expression including the $D=6$ condensate and the dot-dashed (red-wine) is the QCD expression including the $D=8$ contribution. } 
\label{fig:ratiob}
\end{center}
\vspace*{-0.5cm}
\end{figure} 
\nin
 %%%%%%%%%%%%%%%%%%%%%%%%%%%%%%%%%%%%%%%
%\vspace*{-0.25cm}
%%%%%%%%%%%%%%%%%%%%%%%%%%%%%%%%%%%
\begin{center}
%\vspace*{-0.5cm}
{\begin{figure}[hbt]
\begin{center}
{\includegraphics[width=7.cm]{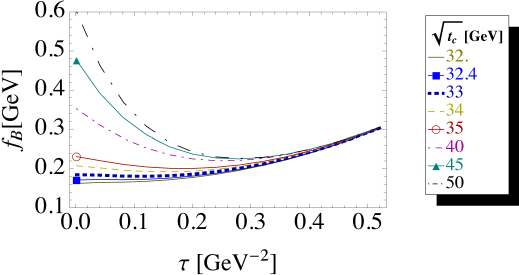}}
\end{center}
%\vspace*{-0.5cm}
%\hspace*{1cm}
%{\includegraphics[width=6cm]{nrcharm.pdf}}
\caption{Behaviour of $f_B$ from Ref.\,\cite{SNFB12} versus the sum rule variable $\tau$ for different values of the QCD continuum threshold $t_c$ and for a given $\mu=$ 3 GeV and $\overline{m}_b(\overline{m}_b) = 4177$ MeV.}
\label{fig:fb}
\vspace*{-0.5cm}
\end{figure}}
\end{center}
%%%%%%%%%%%%%%%%%%%%%%%%%%%%%%%%%%
\vspace*{-0.25cm}
%%%%%%%%%%%%%%%%%%%%%%%%%%%%%%%%%%%
Hopefully, in the examples of harmonic oscillator in quantum mechanics and of the non-relativistic charmonium sum rules, Refs.\,\cite{BELLa,BERTa} have shown that the optimal information 
from the analysis from truncated series is obtained at the minimum or inflexion point in $\tau$ of the approximate theoretical curves (see Figs\,\ref{fig:oscillo1} and \ref{fig:oscillo2} {\it (principle of minimum sensitivity of the physical parameters on the external sum rule variable $\tau$)}.  We illustrate in Figs.\,\ref{fig:ratiob} and \ref{fig:fb} the  analysis for the $\Upsilon$ systems and for the $B$ meson decay constant using relativistic  sum rules. 
%%%%%%%%%%%%%%%%%%%%%%%%
\vspace*{-0.15cm}
\subsection*{\d $t_c$ and $\mu$-stabilities}
\vspace*{-0.1cm}
%%%%%%%%%%%%%%%%%%%%%%%%
Later on, we have extended this $\tau$-stability criterion to the continuum threshold variable $t_c$\,\footnote{In many papers, the optimal value is exracted at the lowest value of $t_c$ ! but the result still increaes with the $t_c$ changes.  } and to the arbitrary Perturbative (PT) subtraction point $\mu$\,\footnote{One can also use the RGE resummed solution which is equivalent to take $\mu^2=1/\tau$ but in some case the value of $\tau$ is relatively large such that the OPE is not well behaved. A such choice of $\mu$ value is often outside the $\mu$-stability region.}. 
The lowest value of $ t_c$ corresponds to the beginning  of $\tau$-stability while the $t_c$-stability 
corresponds to a complete lowest ground state dominance in the spectral integral. In our analysis, we always consider the conservative optimal result inside this $t_c$-region\,\footnote{$\sqrt{t_c}$ is often identified to the mass of the 1st radial excitation which is a crude approximation as the QCD continuum smears all higher state contributions to the spectral function.}. 
%%%%%%%%%%%%%%%%%%%%%%%%
\vspace*{-0.15cm}
\subsection*{\d Ground state versus the QCD continuum}
\vspace*{-0.1cm}
%%%%%%%%%%%%%%%%%%%%%%%%
For some low values of $(t_c,\tau)$, one can have some flat curves ior some (apparent) minimum in $\tau$. Then, to check or/and to restrict the optimal region in this case, one also  requests that the contribution of the ground state to e.g. the spectral integral (e.g. moment sum rule) is larger than the QCD continuum one, which one can formulate in a rigorous way as (see e.g.\,\cite{SNPC} for some examples of applications):
\vspace*{-0.15cm}
\beq
\hspace*{-0.75cm}R_{P/C} \equiv\frac{\int_{t>}^{t_c}dt\,e^{-t\tau}{\,\rm Im}\,\Pi(t)}{\int_{t_c}^{\infty}dt\,e^{-t\tau}{\,\rm Im}\,\Pi(t)}\geq 1.
%\vspace*{-0.15cm}
\eeq
\vspace*{-0.35cm}
%%%%%%%%%%%%%%%%%%%%%%%%%%%%%%%%%
\vspace*{-0.25cm}
\section{Light quark sum rules}
\vspace*{-0.25cm}
%%%%%%%%%%%%%%%%%%%%%%%%%%%%%%%%%
The light quarks sum rules have been used (for the first time) in Refs.\,\cite{BECCHI,SNR} to determine the light quark running masses. Later on, it has been applied to study the light meson properties\,\cite{SNLmeson,SNLREV,RRY}, to extract  the gluon  $\la\alpha_s G^2\ra$, the quark $\la\bar \psi\psi\ra$,  the mixed quark-gluon  $\la\bar \psi G\psi\ra$ and four-quark condensates.  The different determinations  are reviewed in\,\cite{SNB1,SNB2,SNLREV} while the determinations of the light quark masses have been updated in the recent work\,\cite{SNLIGHT}. The results from light quark and $\tau$-decay like sum rules are quoted in Table\,\ref{tab:param}. 

%%%%%%%%%%%%%%%%%%%%%%%%%%%%%%%%%
\vspace*{-0.25cm}
\section{Heavy quark sum rules  for ordinary hadrons}
\vspace*{-0.25cm}
%%%%%%%%%%%%%%%%%%%%%%%%%%%%%%%%%
\d Since the work of SVZ, charmonium and bottomium sum rules have been used to determine the heavy quark mass and the gluon condensates. The analysis has been updated in Ref.\,\cite{SNparam} where (for the first time) the correlations between different parameters ($m_Q,\alpha_s,\la\alpha_sG^2\ra$) have been emphasized using other channels than the vector one. Not taking into account such correlations from some other channels have lead to some apparent discrepancies among some previous determinations. 
%%%%%%%%%%%%%%%%%%%%%%%%%%%%%%%%%%%%%%%
\begin{figure}[hbt]
\begin{center}
%\hspace*{-6cm}{\bf a)}\\
\includegraphics[width=6.5cm]{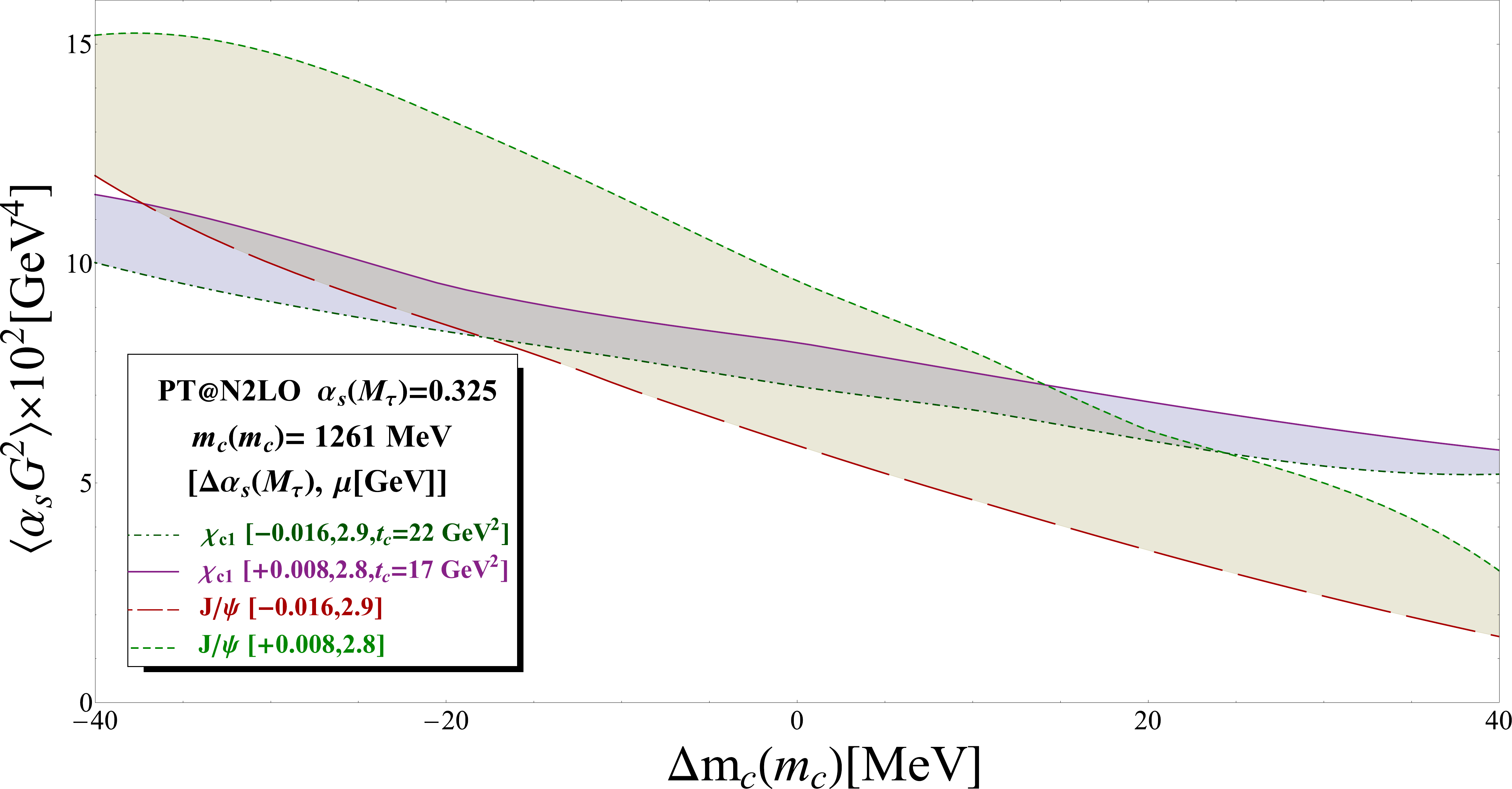}
%\hspace*{-6cm}{\bf b)}\\
%\includegraphics[width=8.cm]{mc-g29.pdf}
\vspace*{-0.5cm}
\caption{\footnotesize  Correlation between $\la\alpha_s G^2\ra$ and $\overline{m}_c(\overline{m}_c)$ for given values of $\alpha_s$ and $\mu$ from $J/\psi$ and $\chi_c$.} 
\label{fig:mc-g2}
\end{center}
\vspace*{-0.5cm}
\end{figure} 
%%%%%%%%%%%%%%%%%%%%%%%%%%%%%%%%%%%%%%%%%

\d This is, e.g., the case of $\la\alpha_sG^2\ra$ from the $J/\psi$ channel where, as one can see in Fig.\,\ref{fig:mc-g2}, one cannot constrain  accurately its value without adding the analysis of the pseudoscalar $\chi_{c1}$ channel. A similar relation between $\bar m_c(\bar m_c)$ and $\la\alpha_sG^2\ra$ from vector channel has been obtained in\,\cite{IOFFEC1} where using, as input, an inappropriate value of $\bar m_c(\bar m_c)$, has lead the authors to a too small value of  $\la\alpha_sG^2\ra$. 
%%%%%%%%%%%%%%%%%%%%%%%%%%%%%%%%%%%%%%%
\begin{figure}[hbt]
\vspace*{-0.3cm}
\begin{center}
\includegraphics[width=6.5cm]{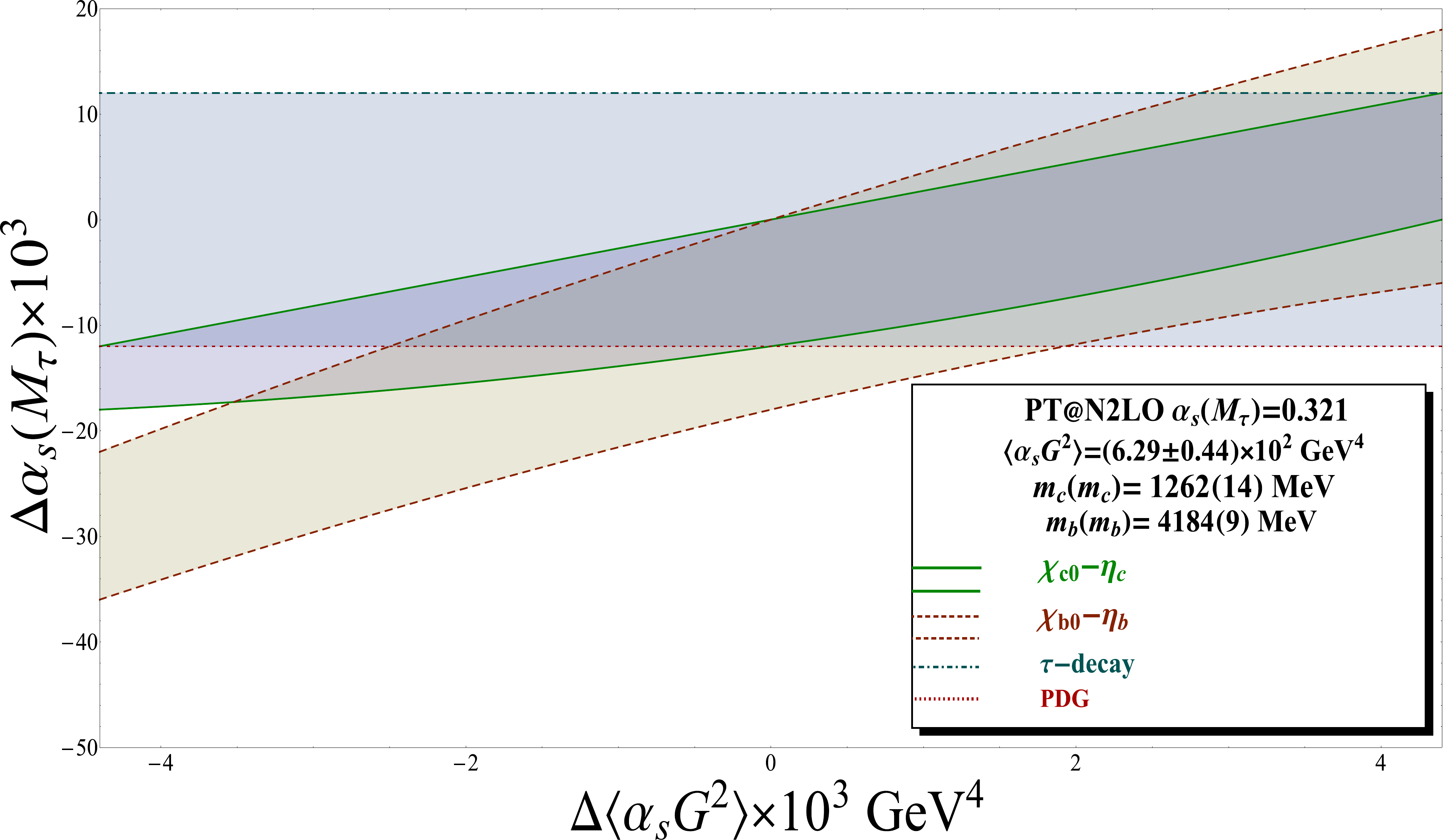}
\vspace*{-0.25cm}
\caption{\footnotesize  Correlation between $\alpha_s$ and $\la \alpha_s G^2\ra$ by requiring that the sum rules reproduce the (pseudo)scalar mass-splittings for given values of $\bar m_{c,b}(\bar m_{c,b})$. The central value of $\la \alpha_sG^2\ra$ in the legend is the one of the previous average obtained in\,\cite{SNparam}.} 
\label{fig:alfas-g2}
\end{center}
\vspace*{-0.75cm}
\end{figure} 
%%%%%%%%%%%%%%%%%%%%%%%%%%%%%%%%%%%%%%
 %%%%%%%%%%%%%%%%%%%%%%%%%%%%%%%%%%%%%%
\begin{figure}[hbt]
\vspace*{-0.5cm}
\begin{center}
\includegraphics[width=7.5cm]{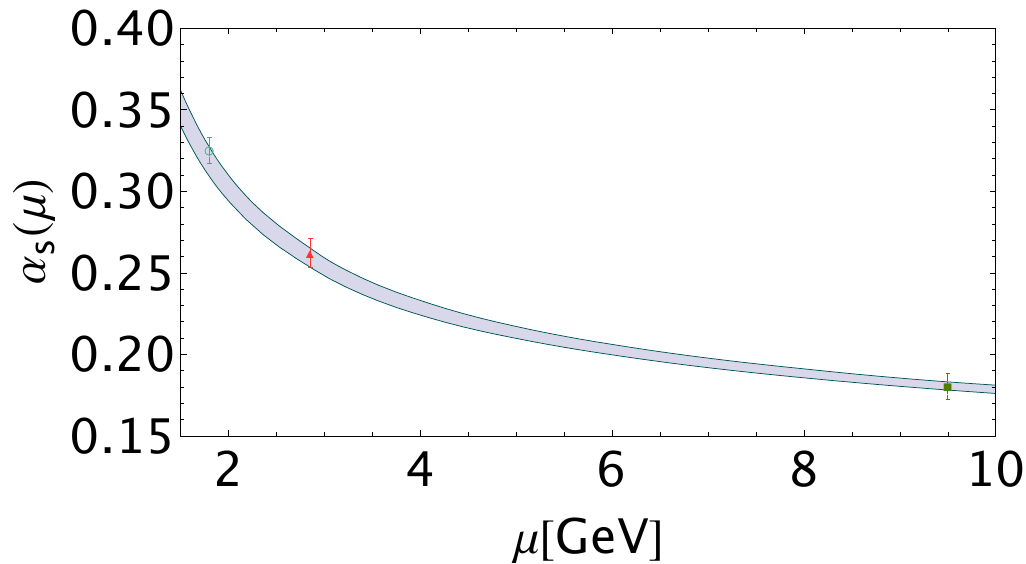}
\vspace*{-0.25cm}
\caption{\footnotesize  Comparison with the running of the world average $\alpha_s(M_Z)=0.1181(11)$\,\cite{BETHKE,PDG} of our predictions at three different scales: $M_\tau$  for the original low moment $\tau$-decay width\,\cite{SNTAU} (open circle),  2.85 GeV for $M_{\chi_{c0}}-M_{\eta_c}$ (full triangle) and  9.5 GeV for $M_{\chi_{b0}}-M_{\eta_b}$ (full square)\,\cite{SNparam}.}
\label{fig:alfas}
\end{center}
\vspace*{-0.75cm}
\end{figure} 
%%%%%%%%%%%%%%%%%%%%%%%%%%%%%%%%%%%%%%%%%%

\d Some other interesting correlations can be found in Ref.\,\cite{SNparam}. 
In particular,  the value of $\alpha_s$ at the subtraction scale $\mu_c=2.85$ GeV and $\mu_b=9.5$ GeV has been determined from the $M_{\chi_{0c(0b)}}-M_{\eta_{c(b)}}$ mass splittings
taking into account its correlation with $\la\alpha_sG^2\ra$ as can be seen in Fig.\,\ref{fig:alfas-g2}:
\vspace*{-0.15cm}
\bea
&&\hspace*{-1cm} \alpha_s(2.85)=0.262(9) \leadsto\alpha_s(M_\tau)=0.318(15)~\nnb\\
&&\hspace*{0.25cm} \leadsto\alpha_s(M_Z)=0.1183(19)(3) ~,\nnb\\
&&\hspace*{-1cm} \alpha_s(9.50)=0.180(8) \leadsto\alpha_s(M_\tau)=0.312(27)\nnb\\
&&\hspace*{0.25cm} \leadsto\alpha_s(M_Z)=0.1175(32)(3) ~.
\label{eq:alfas1}
%\label{eq:glue1}
\eea
%\vspace*{-0.15cm}
The last error is due to the running procedure. 
We have requested that the method reproduces within the errors the experimental mass-splittings by about ($2\sim 3$) MeV. The geometric mean of the two previous values of $\alpha_s$ is\,:
\vspace*{-0.3cm}
\beq
\hspace*{-0.5cm}\alpha_s(M_\tau)=0.317(15) \leadsto\alpha_s(M_Z)=0.1181(19)(3),
\label{eq:alfas}
\eeq
\vspace*{-0.15cm}
which is (surprisingly) in a very good agreement with the world average\,\cite{BETHKE,PDG}\,:
\vspace*{-0.15cm}
\beq
\alpha_s(M_Z)=0.1181(11)~.
\label{eq:asworld}
\eeq
\vspace*{-0.15cm}
and where  one observes a nice 1/Log-behaviour  (see Fig.\,\ref{fig:alfas-g2}) as expected from asymptotic freedom. 
%%%%%%%%%%%%%%%%%%%%%%%%%%%%%%%%%%%%%%%%%%%%%%%
{\scriptsize
\vspace*{-0.35cm}
\begin{table}[hbt]
%\label{tab:param}
 \caption{Values of $\overline{m}_c(\overline{m}_c)$ and $\overline{m}_b(\overline{m}_b)$ in units of MeV coming from most recent QSSR analysis based on stability criteria. Some other determinations can be found in PDG\,\cite{PDG}. }  
%\tbl{
%}
\setlength{\tabcolsep}{0.1pc}
    {\small
  \begin{tabular}{llll}
 % \begin{tabular*}{\textwidth}{@{}l@{\extracolsep{\fill}}llll}
% {\begin{tabular}{@{}llll@{}} \toprule
&\\
\hline
%\hline
%\\
Masses&Values& Sources & Ref.    \\
%\\
\hline
%\\
$\overline{m}_c(\overline{m}_c)$&$1256(30)$  &${J/\psi}$ family&Ratios of LSR\, \cite{SNparam}\\
&$1266(16)$ &$M_{\chi_{0c}-M_{\eta_c}}$&Ratios of LSR\, \cite{SNparam}\\
&1264(6) &${J/\psi}$ family& MOM \& Ratios of MOM\, \cite{SNH18} \\
&1286(66)&$M_D$ & Ratios of LSR\,\cite{SNFB12}\\
&1286(16)&$M_{B_c}$ & Ratios of LSR \cite{SNBc2} \\
%\\
&{\it 1266(6)}&{\it Average}& \cite{SNBc2} \\
%\\
%\hline
%\\
$\overline{m}_b(\overline{m}_b)$&$4192(17)$&${\Upsilon}$ family&Ratios of LSR\,\cite{SNparam}\\
&4188(8) &$\Upsilon$ family &MOM \& Ratios of MOM\, \cite{SNH18} \\
&4236(69) &$M_B$&Ratios of MOM \& of LSR\,\cite{SNFB12}\\
&4213(59)&$M_B$& Ratio of HQET-LSR\,\cite{SNHQET13}\\
&4202(7)&$M_{B_c}$ & Ratios of LSR\,\cite{SNBc2}\\
%\\
&{\it 4196(8)}&{\it  Average}& \cite{SNBc2}\\
%\\
\hline
%\hline
\end{tabular}
}
\label{tab:hmass}
%\caption{%\scriptsize   
\end{table}
} 
%%%%%%%%%%%%%%%%%%%%%%%%%%%%%%
\vspace*{-0.25cm}
%%%%%%%%%%%%%%%%%%%%%%%%%%%%%%%%%%%%%%%%%%
{\scriptsize
\begin{table}[hbt]
 \caption{QCD parameters from light  and  heavy quarks QSSR (Moments, LSR and ratios of sum rules) within stability criteria. The running light quark masses and condensates have been evaluated at 2 GeV within the SVZ expansion without instantons contributions disfavoured by lattice results\,\cite{LATTLIGHT}.}
%%%%%%%%%%%%%%%%%%%%%%%%%%%%%%%%%%%%%%%%%%
\setlength{\tabcolsep}{0.0pc}
    {\small
  \begin{tabular}{llcc}
% {\begin{tabular}{@{}lll@{}} \toprule
&\\
\hline
%\hline
%\\
Parameters&Values&Sources& Ref.    \\
%\\
\hline
\it Heavy \\
$\alpha_s(M_Z)$& $0.1181(16)(3)$&$M_{\chi_{0c,b}-M_{\eta_{c,b}}}$&
%\cite{SNTAU,BNPa,BNPb,BETHKE,PDG}\\
\cite{SNparam} \\
$\overline{m}_c(\overline{m}_c)$ [MeV]&$1266(6)$ &$D, B_c \oplus$&(see Table\,\ref{tab:hmass}) \\
% \cite{SNH10a,SNH10b,SNH10c,PDG,IOFFEa,IOFFEb}\\
%\cite{SNm20,SNparam,SNbc20,SNmom18,SNFB15}\\
&&$ {J/\psi}, \chi_{c1},\eta_{c}$ &\\
$\overline{m}_b(\overline{m}_b)$ [MeV]&$4196(8)$ &$B,B_c\oplus{\Upsilon}$&  (see Table\,\ref{tab:hmass}) \\
%\cite{SNH10,SNH11,SNH12,SNparam,SNbc20,SNFB15,SNmom18}\\
%\cite{SNbc20,SNmom18}\\
$\la\alpha_s G^2\ra$ [GeV$^4$]& $6.49(35) 10^{-2}$&Light, Heavy &
%\cite{SNTAU,LNT,SNIa,SNIb,YNDU,BELLa,BELLb,BELLc,SNH10a,SNH10b,SNH10c,SNG1,SNG2,SNGH}\\
 \cite{SNparam,SNREV1}\\
${\la g^3  G^3\ra}/{\la\alpha_s G^2\ra}$& $8.2(1.0)$[GeV$^2$]&${J/\psi}$&
\cite{SNcb1}\\
%\\
 \it{Light} \\
$\hat \mu_\psi$ [MeV]&$253(6)$ &Light &\,\cite{SNB1,SNB2,SNp15,SNLIGHT} \\ %\,\cite{SNB1,SNB2,SNp15,SNLIGHT} \\
$\la\overline{\bar \psi\psi}\ra(2)$ [MeV]$^3$&$-(276\pm 7)^3$ &--&-- \\
$\kappa\equiv\la \bar ss\ra/\la\bar dd\ra$& $0.74(6)$&Light, Heavy&\cite{SNB1,SNB2,SNp15,SNLIGHT,HBARYON1,HBARYON2}\\
$\hat m_u$ [MeV]&$3.05\pm 0.32$&Light &\,\cite{SNB1,SNB2,SNp15,SNLIGHT} \\
$\hat m_d$ [MeV]&$6.10\pm0.57$ &-- &-- \\
$\hat m_s$ [MeV]&$114(6)$ &-- & -- \\
$\overline{ m}_u$ (2) [MeV]&$2.64\pm 0.28$ &-- &-- \\
$\overline{ m}_d$ (2) [MeV]&$5.27\pm 0.49$ &-- & -- \\
$\overline{ m}_s$ (2) [MeV]&$98.5\pm 5.5$ &-- & --\\
%$\la \bar dd\ra(2) $&$-(275.7\pm 6.6)^3$ MeV$^3$&\cite{SNB1,SNmass}\\
$M_0^2$ [GeV$^2$]&$0.8(2)$ &Light, Heavy&\,\cite{SNB1,SNB2,IOFFE,DOSCH,PIVOm,SNhl}\\

%\cite{SNH10} \\
$\rho \alpha_s\la \bar \psi\psi\ra^2\times 10^{4}$ &$5.8(9) $[GeV$^6$] &Light, $\tau$&\cite{FESR2,SNTAU,LNT,DOSCH}\\
%$\hat m_s$&$(0.114\pm0.006)$ GeV &\cite{SNB1,SNTAU9,SNmassa,SNmassb,SNmass98a,SNmass98b,SNLIGHT}\\
%$\kappa\equiv \la \bar ss\ra/\la\bar dd\ra$& $(0.74^{+0.34}_{- 0.12})$&\cite{HBARYONa,HBARYONb,SNB1}\\
\hline
%\hline
\end{tabular}}

\label{tab:param}
%}
%\caption{%\scriptsize   
\vspace*{-0.25cm}
\end{table}
} 
%%%%%%%%%%%%%%%%%%%%%%%%%%%%%%
%%%%%%%%%%%%%%%%%%%%%%%%%%%%%%%%%%
\vspace*{-0.25cm}
{\scriptsize
\begin{center}
\begin{table}[hbt]
 \caption{\scriptsize  Heavy-light decay constants $f_H$ within $\mu$ and $t_c$-stability at N2LO. $f_H$ is normalized as $f_\pi=132$ MeV.}
 \label{tab:decay}
 \setlength{\tabcolsep}{0.8pc}
\begin{tabular}{lllc}
\hline 
\footnotesize  Channel  &\footnotesize Values [MeV]&\footnotesize Bounds [MeV]&\footnotesize $f_{H_s}/f_H$\\
\hline
\footnotesize $D$ &\footnotesize 204(6)&\footnotesize$\leq 218(2)$& \footnotesize 1.170(23) \\
\footnotesize ${D_s}$ &\footnotesize 243(5)&\footnotesize$\leq 254(2)$& -- \\
\footnotesize $B$ &\footnotesize 204(5)&\footnotesize$\leq 235(4)$&\footnotesize 1.154(21) \\
\footnotesize ${B_s}$ &\footnotesize 235(4)&\footnotesize$\leq 251(6)$& -- \\
\footnotesize ${D^*}$ &\footnotesize 250(8)&\footnotesize$\leq 266(8)$&\footnotesize 1.215(30) \\
\footnotesize ${D^*_s}$ &\footnotesize 290(11)&\footnotesize$\leq 287(18)$&-- \\
\footnotesize ${B^*}$ &\footnotesize 210(6)&\footnotesize$\leq 295(15)$&\footnotesize 1.020(11) \\
\footnotesize ${B^*_s}$ &\footnotesize 221(7)&\footnotesize$\leq 317(17)$&-- \\
%\footnotesize ${B_c}$ &\footnotesize 436(40)&\footnotesize$\leq 466(16)$&-- \\
\footnotesize ${D^*_0}$ &\footnotesize 220(11)&&\footnotesize 0.922(15) \\
\footnotesize ${D^*_{s0}}$ &\footnotesize 202(15)&& -- \\
\footnotesize ${B^*_0}$ &\footnotesize 278(12)&&\footnotesize 1.064(10)\\
\footnotesize ${B^*_{s0}}$ &\footnotesize 255(15)&& -- \\

%&&\\
\hline
%\hline
\end{tabular}
\end{table}
\end{center}
 }
\vspace*{-0.75cm}
\nin
%%%%%%%%%%%%%%%%%%%%%%%%%%%%%%%%%%%%%%%%%%%%%%

\d As mentioned earlier in Section\,\ref{sec:ope}, the ratio $\la g^3 G^3\ra/\la\alpha_sG^2\ra$ condensate has been also extracted (for the first time) from charmonium sum rules where a  large deviation from the dilute gas instanton estimate has been observed. 

\d These determinations of the QCD parameters from heavy quark sum rules have been reviewed recently in Ref.\,\cite{SNREV1} and summarized in the Tables\,\ref{tab:hmass} and \ref{tab:param}. 

%%%%%%%%%%%%%%%%%%%%%%%%%%%%%%%%%
 \vspace*{-0.35cm}
\section{Heavy-Light quark sum rules}
\vspace*{-0.25cm}
%%%%%%%%%%%%%%%%%%%%%%%%%%%%%%%%%

%%%%%%%%%%%%%%%%%%%%%%%%%%
\subsection*{\d $D,B$-like mesons}
\vspace*{-0.1cm}
%%%%%%%%%%%%%%%%%%%%%%%%%%%%%%%%%
In this channel, the sum rules have been used to extract the decay constants of the $D,D^*$ and of their strange partners $D_s,D^*_s$ mesons, their  chiral partners and their $B$-like analogue.  The masses of the observed $D,D^*$ mesons and their $B$-like analogue have been used to determine the running $\overline m_c(\overline m_c)$ and  $\overline m_b(\overline m_b)$ quark masses (see Table\,\ref{tab:hmass}). The average values of the decay constants and the SU(3) breaking ratios come from\,\cite{SNbc15} and references therein. The results are compiled in Table\,\ref{tab:B-coupling}.
%%%%%%%%%%%%%%%%%%%%%%%%%%%%%%%%%
\vspace*{-0.25cm}
\subsection*{\d $B_c$-like mesons}
\vspace*{-0.15cm}
%%%%%%%%%%%%%%%%%%%%%%%%%%%%%%%%%
Similar analysis has been done for the $B_c$-like mesons. 

\begin{figure}[hbt]
\begin{center}
%\hspace*{-6cm}{\bf a)}\\
\includegraphics[width=7.5cm]{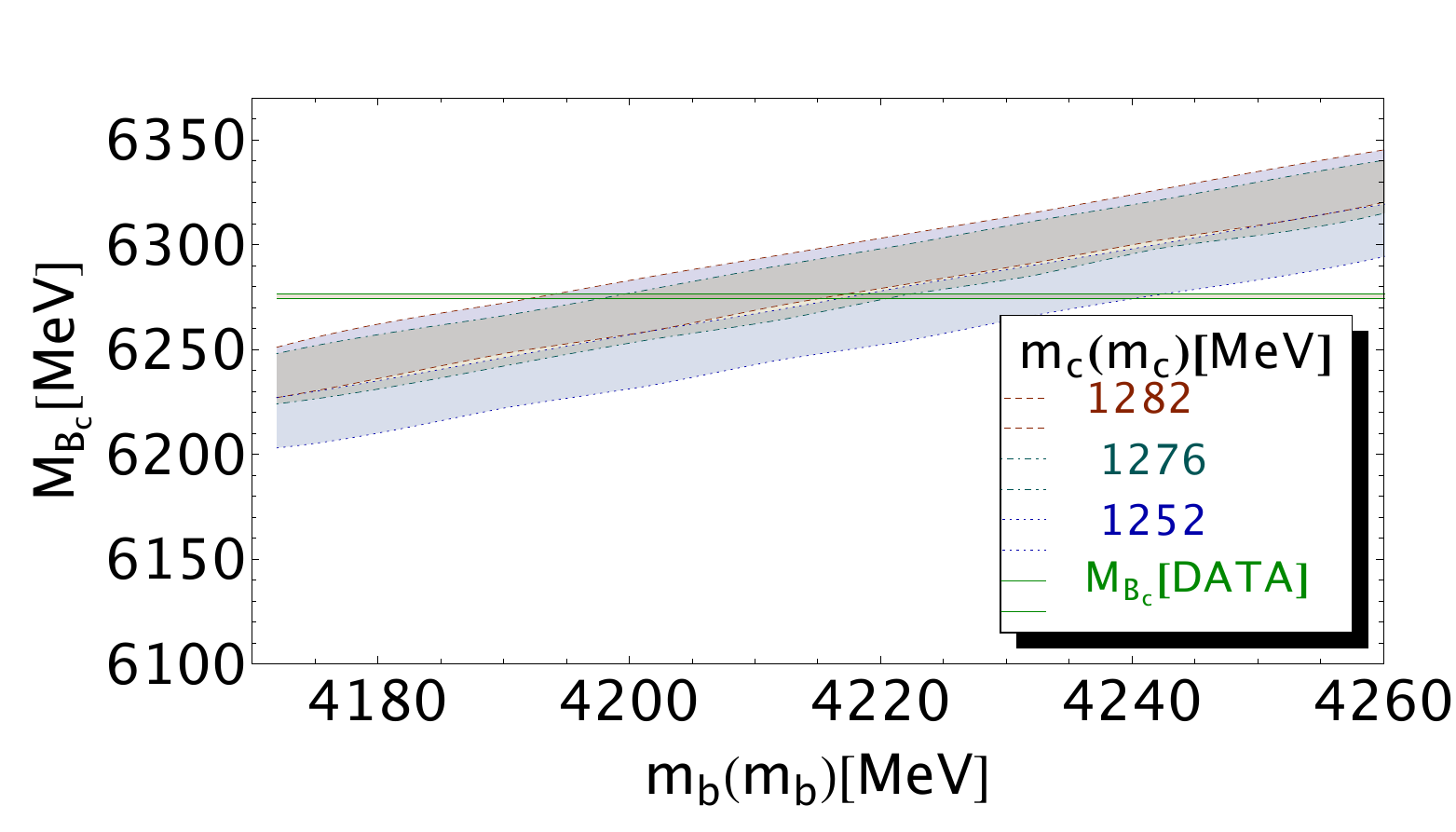}
%\hspace*{-6cm}{\bf b)}\\
%\includegraphics[width=8.cm]{mc-g29.pdf}
%\vspace*{-0.5cm}
 \caption{\footnotesize $M_{B_c}$ as function of $\overline m_b(\overline m_b)$ for different values of $\overline m_c(\overline m_c)$,  at the stability point  $\mu$=7.5 GeV and for the range  of $\tau$-stability values $\tau=(0.30-0.32)$ GeV$^{-2}$. }
\label{fig:mc-mb}
\end{center}
\vspace*{-0.25cm}
\end{figure}

The $B_c$ mass prediction taking into account the correlation between $\overline m_c(\overline m_c)$ and  $\overline m_b(\overline m_b)$  is  given in Fig.\,\ref{fig:mc-mb}. The resulting values of $\overline m_c(\overline m_c)$ and  $\overline m_b(\overline m_b)$ are quoted in Table\,\ref{tab:hmass}. 
%%%%%%%%%%%%%%%%%%%%%%%%%%%%%%%%%%%%%%%
%\vspace*{-0.5cm}
%%%%%%%%%%%%%%%%%%%%%%%%%%%%%%%%%%%%%%%%%

%%%%%%%%%%%%%%%%%%%%%%%%%%%%%%%%%%%%%%%%
%\vspace*{-0.5cm}
{\scriptsize
\begin{table}[hbt]
 \caption{Values of the masses from LSR and HQS compared with lattice  and potential models (PM) results.}
\setlength{\tabcolsep}{0.12pc}
    {\small
  \begin{tabular}{ccccc}
 % \begin{tabular*}{\textwidth}{@{}l@{\extracolsep{\fill}}llll}
% {\begin{tabular}{@{}llll@{}} \toprule
&\\
\hline
%\hline
%\\
Channel&LSR& HQS &Lattice\,\cite{LATTBC}&PM\,\cite{QUIGG}  \\
%\\
\hline
%\\
Masses \\
$B^*_c(1^{--})$&$6451(86)$  &6315(1)&6331(7)&6330(20)\,\cite{BAGAN}\\
$B^*_{0c}(0^{++})$&6689(198)&6723(29)&6712(19)&6693 \\
$B_{1}(1^{++})$&6794(128)&6730(8)&6736(18)&6731 \\
$B_{c2}(2^{++})$&--&6741(8)&--& 7007\\
%&\\
\hline 
%&\\
$B^*_{0}(0^{++})$&5701(196)\,\cite{SNhl05}&5733& \\
\hline
%\hline
\end{tabular}
}
\label{tab:B-coupling}
\vspace*{-0.25cm}
%\caption{%\scriptsize   
\end{table}
} 
% \vspace*{-0.5cm}
%\vspace*{-0.5cm}
%%%%%%%%%%%%%%%%%%%%%%%%%%%%%%
{\scriptsize
\begin{table}[hbt]
%\label{tab:param}
 \caption{Values of the decay constants $f_H$ in units of MeV using as input the values of the masses from LSR and HQS quoted in Table\,\ref{tab:hmass}. }  
 
\setlength{\tabcolsep}{0.12pc}
    {\small
  \begin{tabular}{ccccc|c}
 % \begin{tabular*}{\textwidth}{@{}l@{\extracolsep{\fill}}llll}
% {\begin{tabular}{@{}llll@{}} \toprule
%&\\
\hline
%\hline
%\\
Masses&$B_c(0^{--})$&$B^*_c(1^{--})$ &$B^*_{0c}(0^{++})$&$B_{1c}(1^{++})$&$B^*_{0}(0^{++})$\\
%\\
\hline
LSR&371(17)\,\cite{SNbc20}&442(44)&155(17)&274(23)&-- \\
HQS&--&387(15)&158(9)&266(14)&271(26)\,\cite{SNbc15} \\

\hline
%\hline
\end{tabular}
}
\label{tab:hdecay}
%\caption{%\scriptsize   
\end{table}
} 
%%%%%%%%%%%%%%%%%%%%%%%%%%%%%%%%%
The masses and couplings of the $B_c$-like mesons from LSR\,\cite{SNbc20b} are given in Table\,\ref{tab:hmass} and \ref{tab:hdecay} which are compared with the Heavy Quark Symmetry\,\cite{SNbc20b}, Potential models and Lattice calculations. An upper bound for the the $B_c(2S)$ coupling  is also derived\,\cite{SNbc20}:
%\vspace*{-0.15cm}
\beq
f_{B_c}(2S ) \leq (139 \pm  6) ~{\rm MeV}.
\eeq
\vspace*{-0.15cm}
%\vfill\eject
%%%%%%%%%%%%%%%%%%%%%%%%%%%%%%%%%
\vspace*{-0.25cm}
\section{Gluonia / Glueballs sum rules}
\vspace*{-0.15cm}
%%%%%%%%%%%%%%%%%%%%%%%%%%%%%%%%%
\d NSVZ\,\cite{NSVZ}  have also used the sum rules to predict qualitatively the gluonia scale which has been recently revisited in\,\cite{CNZa} including the tachyonic gluon mass contribution. This effect has restored the large value of some scales found by NSVZ.  Since then, some progress has been done for more quantitative predictions of the spectra. 
%%%%%%%%%%%%%%%%%%%%%%%%%%%%%%
\vspace*{-0.25cm}
\subsection*{\d Pseudoscalar digluonium  and the proton spin}
%%%%%%%%%%%%%%%%%%%%%%%%%%%%%%
This channel is described by the two-point function:
\vspace*{-0.15cm}
\beq
\psi_P(q^2) =\ga{8\pi}\dr^2i\int d^4x \,e^{iqx}\la 0 \vert Q(x)Q^\dagger(0)\vert 0\ra
\eeq
\vspace*{-0.15cm}
 associated to the divergence of $U(1)_A$ axial current which reads for $n_f$ flavours:
 \vspace*{-0.15cm}
 \beq
 \partial_\mu J^\mu_5(x)= \sum_{i=u,d,s}2m_i\bar\psi_i\psi_i+2n_f Q(x)~, 
 \eeq
 \vspace*{-0.25cm}
 where :
 \vspace*{-0.15cm}
\beq
Q(x)\equiv
 \ga\frac{\alpha_s}{16\pi}\dr\epsilon_{\mu\nu\rho\sigma}G^{\mu\nu}_a(x)G^{\rho\sigma}_a(x),
\eeq
\vspace*{-0.15cm}
  $a=1,...,8$ is the colour index. In earlier works\,\cite{SNU1,SNG1}, this sum rule has been used  to extract the gluon component of the $\eta'$-mass and decay constant, 
  the topological charge $\chi(0)\equiv \psi_P(0)/(8\pi)$ and its slope $\chi'(0)$. A recent update including N2LO  PT corrections lead, in the chiral limit, to (in units of MeV) \,\cite{SNGP22} :
  \vspace*{-0.15cm}
 \beq
 f_{\eta_1}=905(72), ~~~~~~ \sqrt{\chi'(0)\vert_{chiral}}=24.3(3.4)~,
 \eeq
 \vspace*{-0.15cm}
 compared to the one at NLO\,\cite{SHORE} and the one from pure Yang-Mills\,\cite{SNG1}: $ {\chi'(0)\vert_{YM}}=-(7\pm 3)~{\rm MeV}$. This result is also smaller than the OZI value:  $\sqrt{\chi'(0)\vert_{\rm OZI}}=f_\pi/\sqrt{6}=38$ MeV\,\footnote{ $f_{\eta1}$ is normalized as $f_\pi=$ 93 MeV like other gluonia decay constants in this section of gluonia.}. Used in the relation with the proton singlet form factor $G_A(0)(Q^2)$ appearing in the first
moment of the polarised proton structure function $g_1^P$\,\cite{SHORE1,SHORE2} (Ellis-Jaffe sum rule\,\cite{JAFFE}), one obtains at N2LO\,\cite{SNGP22}:
\vspace*{-0.25cm}
 \bea
\hspace*{-0.5cm}  G_A^{(0)}\vert_{\rm LSR}(Q^2&=&10~\rm GeV^2)= G_A^{(0)}\vert_{\rm OZI} 
  \frac{\sqrt{{\chi'(0)}}\vert_{\rm LSR}}
 {\sqrt{{\chi'(0)}}\vert_{\rm OZI}}\nnb\\
 &=& (0.340\pm 0.050),
%  \equiv \Delta\Sigma=\Delta u + \Delta d + \Delta s=3F-D=0.579\pm 0.021~,
\label{eq:ga}
\eea
%\vspace*{-0.15cm}
after running $\chi'(0)\vert_{\rm LSR}$ from 2 to 10 GeV$^2$ where $G_A^{(0)}\vert_{\rm OZI} =0.579\pm 0.021$. As a result, the first moment of the polarized proton structure function reads:
 \bea
 \vspace*{-0.15cm}
\Gamma^p_1(Q^2=10 ~\rm GeV^2)&\equiv& \int_0^1 dx\,g_1^P(x,Q^2)\nnb\\
&=& (0.144\pm 0.005),
\label{eq:spin1}
\vspace*{-0.15cm}
\eea
in excellent agreement with the world average $(0.145\pm 0.014)$\,\cite{SMC2} and the recent COMPASS\,\cite{COMPASS} and HERMES data\,\cite{HERMES}. 
 %%%%%%%%%%%%%%%%%%%%%%%%%%%%%%
\subsection*{\d Pseudoscalar digluonia spectrum}
\vspace*{-0.0cm}
%%%%%%%%%%%%%%%%%%%%%%%%%%%%%%
Using a one resonance parametrization of the spectral function, one obtains from ${\cal L}_0$ and ${\cal R}_{10}$\,\cite{SNG0}:
\vspace*{-0.15cm}
\beq
\hspace*{-0.5cm}  f_P\simeq (8\sim 12)~{\rm MeV},~~M_P\simeq (2.05\pm 0.10)~{\rm GeV}, 
 \eeq
 \vspace*{-0.15cm}
 and the upper bound from the positivity of the spectral function:
 \vspace*{-0.15cm}
 \beq
 M_P\simeq (2.34\pm 0.42)~{\rm GeV}.
 \eeq
 \vspace*{-0.15cm}
 The mass prediction is comparable with the ones from other sum rules determinations\,\cite{ASNER,FORKEL,ZHUG}.
 
Parametrizing the spectral function beyond the one resonance model, a recent
analysis shows that there can be more states above  the $\eta'$ (see Table 2 of
\cite{SNGP22}):
\vspace*{-0.15cm}
\beq
\hspace*{-0.5cm} P_1(1397\pm 81),~ P'_1(1541\pm 118)~, P_2(2751\pm 140),
\eeq
\vspace*{-0.15cm}
 where the highest mass is  comparable with  the lattice value \,\cite{RAGO}. The corresponding couplings are:
 \vspace*{-0.15cm}
 \beq
\hspace*{-0.5cm} f_{P_1}=594(144),~f_{P'_1}=205(282),~f_{P_2}=500(42). 
 \eeq
 \vspace*{-0.15cm}
 $ P'_1$ is expected to the the radial excitation of $P_1$ which is (a posteriori) justified by its weak coupling to the current via its decay constant. These unmixed states may explain the nature of the observed $\eta(1405), \eta(1495)$ and $\eta(1760)$ while the $P_2(2751)$ remains to be discovered. 
% %%%%%%%%%%%%%%%%%%%%%%%%%%%%%%
 \vspace*{-0.25cm}
\subsection*{\d Scalar digluonia spectra}
\vspace*{-0.cm}
%%%%%%%%%%%%%%%%%%%%%%%%%%%%%%
Similar analysis has been done in the scalar channel by working with the two-point function: 
\vspace*{-0.15cm}
\beq
\hspace*{-0.5cm}\psi_G(q^2)=16i\int d^4x \,e^{iqx}\la 0 \vert \ga\theta^\mu_\mu\dr_G(x) \ga\theta^\mu_\mu\dr_G^\dagger(0)\vert 0\ra
\eeq
\vspace*{-0.15cm}
built from the gluon component of the trace of the energy-momentum tensor:
\vspace*{-0.15cm}
\bea
\hspace*{-0.5cm} \theta^\mu_\mu&=&\frac{1}{4} \beta(\alpha_s) G^{\mu\nu}_aG^a_{\mu\nu}+\ga1+\gamma_m\dr\sum_{u,d,s}m_i\bar\psi_i\psi_i ,
\label{eq:theta}
\eea
%\vspace*{-0.1cm}
%\vspace*{-0.25cm}
with :  $\gamma_m=2\alpha_s/\pi+\cdots$ is the quark mass anomalous dimension and $\beta(\alpha_s)$ is the $\beta$-function. Working with the substracted ${\cal L}_0$  and unsubtracted sum rule ${\cal L}_{-1}$ leads to inconsistencies  for a ``one resonance" parametrization of the spectral function as the two sum rules stabilize at very different values of $\tau$ due to the strong effect of the subtraction constant $\psi_G(0)$ to ${\cal L}_{-1}$\,\cite{NSVZ} which pushes its stabilty to lower values of $\tau$ making it less sensitive to the low mass hadron contrary to usual expectations. Starting the analysis with the unsubtracted high degree of moments ${\cal R}_{21}^c$, and ${\cal L}_{1,2}^c$, one obtains\,\cite{SNG0,VENEZIA}:
\vspace*{-0.1cm}
\beq
\hspace*{-0.5cm} M_G=(1.50\pm 0.19)~{\rm GeV}~, ~~~f_G=(390\pm 145)~{\rm MeV},
\eeq
\vspace*{-0.1cm}
and :
%\vspace*{-0.25cm}
%\beq
$
M_G\leq (2.16\pm 0.16)~{\rm GeV},
$
%\eeq
after using positivity at the minimum of ${\cal R}_{21}^c$.  Using this result into ${\cal L}_{-1,0}$ within a ``two resonances" parametrization of the spectral function,
one obtains:
\vspace*{-0.15cm}
\beq
f{\sigma_B}\approx1.0~{\rm GeV},~~~M{\sigma_B}\approx 1.0~{\rm GeV},
\eeq
\vspace*{-0.15cm}
where we have used\,\cite{NSVZ}:
\vspace*{-0.15cm}
\beq
\hspace*{-0.5cm}\psi_G(0)\vert_{\rm LET}\simeq -\frac{16}{\pi} \beta_1\la\alpha_s G^2\ra=(1.46\pm 0.08)~{\rm GeV}^4,
\label{eq:let}
\eeq
\vspace*{-0.15cm}
from low-energy theorem (LET) to be checked later on. We use  the previous result into the vertex function :
\vspace*{-0.15cm}
\beq
V(q^2)\equiv \la\pi\vert \theta^\mu_\mu \vert\pi \ra =\int_0^\infty \frac{dt}{t-q^2-i\epsilon}\frac{1}{\pi}{\rm Im} V(t),
\eeq
with $q\equiv p_1-p_2$ where $V(0)=2m_\pi^2$. Using the fact that $V(0)=0$ in the chiral limit and $V'(0)=1$, 
one obtains, by saturating the spectral function by the $\sigma_B$ and its radial excitation $\sigma'_B$, the low-energy sum rules:
\vspace*{-0.15cm}
\beq
\hspace*{-0.7cm}\sum_{S\equiv \sigma_B,\sigma'_B} g_{S\pi^+\pi^-}\sqrt{2}f_S\simeq 0, 
 \sum_{S\equiv \sigma_B,\sigma'_B} g_{S\pi^+\pi^-}\frac{\sqrt{2}}{4}\frac{f_S}{M_G^2}\simeq 1,~
\eeq
\vspace*{-0.15cm}
where $g_{S\pi^+\pi^-}$ is the $S\pi^+\pi^-$ coupling normalized as:
\vspace*{-0.15cm}
\beq
\hspace*{-0.5cm}\Gamma [\sigma_B\to \pi^+\pi^-+2\pi^0]=\frac{3}{2}\frac{\vert g_{\sigma_B\pi^+\pi^-} \vert^2}{16\pi M_{\sigma_B}}\ga 1-\frac{4m_\pi^2}{M^2_{\sigma_B}}\dr^{1/2}.
\label{eq:sigwidth} 
\eeq
\vspace*{-0.15cm}
Fixing for definiteness $M_{\sigma'}=1.37$ GeV, one deduces  ${\rm for}~M_{\sigma_B} \simeq 1.07 ~{\rm GeV}$:
\beq
\hspace*{-0.25cm}\Gamma(\sigma_B\to\pi^+\pi^-+\pi^0\pi^0)\simeq 873~{\rm MeV}. 
\eeq

This result has motivated the interpretation that the broad $\sigma/ f_0(500)$ can be a good candidate for a low mass scalar gluonium\,\cite{SNG0,VENEZIA}. Different analysis of the $\gamma\gamma$ and $\pi\pi$ scatterings data\,\cite{MNO,OCHS,SNGS22} have confirmed this result where, within a Breit-Wigner or On-shell  parametrization of the data,  the $\sigma(500)$, in the complex plane, becomes a 920 MeV resonance with a width of about 700 MeV in the real axis. The presence of a low mass  glueball state around (0.6-1.25) GeV are expected from some other sum rules analysis\,\cite{STEELESG,FORKEL}, in the Dragon model of\,\cite{MINK,OCHS} and from a strong coupling analysis of the gluon propagator\,\cite{FRASCA}. A double resonance around 0.8 and 1.6 GeV is also found from a Gaussian sum rule where the lighter has a large width \,\cite{STEELESG} and from dispersive approach\,\cite{HSIANG}. 

However, though we have found two gluonia candidates :  $\sigma_B(1.0)$ and G(1.5), these two states are not sufficient to explain the numerous $I=0$ scalar states below 2 GeV.  In Ref.\,\cite{SNGS22}, we have extended the analysis by parametrizing the spectral function beyond ``two resonances" model and using various higher moments. In this way, we  obtain in units of MeV:
\vspace*{-0.15cm}
\bea
M_{\sigma_B}=1070(126) , ~~M_{G_1}=(1548\pm 121),\nnb\\
f_{\sigma_B}=456(157), ~~f_{G_1}=365(110),
\eea
\vspace*{-0.15cm}
and the radial excitations :
%\vspace*{-0.15cm}
\bea
M_{\sigma'_B}&=&1110(117), ~M_{G'_1}=1563(141),\nnb\\
M_{G_2}&=&2992(221).
\eea
%\vspace*{-0.15cm}
Their effective coupling which might be the sum of some other higher mass radial excitations are:
\vspace*{-0.15cm}
\bea
f^{eff}_{\sigma'_B}&=&648(216), ~f^{eff}_{G'_1}=1000(230)\nnb\\
f^{eff}_{G_2}&=&797(74).
\eea
%\vspace*{-0.15cm}
One can compare $f^{eff}_{\sigma'_B}$ with the one $f_{\sigma'(1370) }=329(30)$ MeV obtained from low-energy vertex sum rule. 

One can observe the one to one correspondence of these states with their chiral pseudoscalar partners obtained in the previous section. One expects that the observed $\sigma(500),~f_0(980),~f_0(1370),~f_0(1500),~f_0(1710)$ emerge from a mixing among these gluonia  or/and with quarkonia states, while the large width of $\sigma_B\to \pi\pi$ is due to the OZI violation in this channel. 
 %%%%%%%%%%%%%%%%%%%%%%%%%%%%%%
 \vspace*{-0.35cm}
\subsection*{\d Conformal charge and its slope}
\vspace*{-0.15cm}
%%%%%%%%%%%%%%%%%%%%%%%%%%%%%%
The moments ${\cal L}_{-1}^c$ and ${\cal L}_0^c$ have been also used to extract the conformal charge $\psi_G(0)$ in order to check the LET result in Eq.\,\ref{eq:let}
while its slope $\psi'_G(0)$ has been extracted from ${\cal L}_2c\oplus  \psi_G(0)$ 
One obtains:
 \vspace*{-0.15cm}
\bea
\psi_G(0)&=&(2.09\pm 0.29)~{\rm GeV}^4,\nnb\\
\psi'_G(0)&=&-(0.2\pm 0.3)~{\rm GeV}^2~,
\eea
 \vspace*{-0.3cm}
compared to the LET estimate in Eq.\,\ref{eq:let}.  
% %%%%%%%%%%%%%%%%%%%%%%%%%%%%%%
% \vspace*{-0.15cm}
\subsection*{\d $2^{++}$ Tensor digluonium}
\vspace*{-0.15cm}
%%%%%%%%%%%%%%%%%%%%%%%%%%%%%%
This channel is described by the two-point function associated to the 
gluon component of the energy momentum tensor:
\vspace*{-0.25cm}
\beq
\theta^g_{\mu\nu}=-G^\rho_\mu G_{\rho\nu}+\frac{1}{4}g_{\mu\nu}G_{\rho\sigma}G^{\rho\sigma}
\vspace*{-0.3cm}
\eeq
Contrary to the case of (pseudo)scalar channels, it cannot be affected by the eventual contribution due to instantons. However, in Ref.\,\cite{CNZa}, it has been shown that the eventual contribution  of the tachyonic gluon mass in the OPE restores the universality of the gluonium scales for diffrent channels.  Its QCD expression has been obtained by Ref.\,\cite{NSVZ} to LO.  Within a ``one resonance" parametrization, one obtains to LO\,\cite{SNG0}:
\vspace*{-0.1cm}
\beq
M_T=(2.0\pm 0.1)~{\rm GeV},~~f_T\simeq (80\pm 14)~{\rm MeV}.
\eeq
\vspace*{-0.55cm}
%%%%%%%%%%%%%%%%%%%%%%%%%%%%%%%%%
\subsection*{\d Trigluonium sum rules}
\vspace*{-0.15cm}
%%%%%%%%%%%%%%%%%%%%%%%%%%%%%%%%5
Since the pioneer work of Ref.\,\cite{PABAN} from the analysis of the two-point function of the scalar  $g^3f_{abc}G^aG^bG^c$ trigluonium current which predicts.\,\cite{PABAN,SNB2} :
 \vspace*{-0.15cm}
\beq
M_{G3}\simeq 3.1~{\rm GeV},~~~f_{G3}\simeq 62~{\rm MeV},
\eeq
 \vspace*{-0.15cm}
and where the digluonium-trigluonium scalar mixing is tiny $(\theta\simeq 4^0)$, there are new sum rules applications in different channels\,\cite{PIMIKOV,QIAO,CHEN,CHEN2,SNG3}. 

Though there are some technical points to be solved among different results (check of the QCD expressions and of the regions for extracting the optimal results), the main feature is that these trigluonium bound states are  higher than the digluonia analogue by about (1--2) GeV which is intuitively expected for high dimension operator currents. 
%%%%%%%%%%%%%%%%%%%%%%%%%%%%%%%%%
\vspace*{-0.15cm}
\section{Hybrid mesons sum rules}
\vspace*{-0.15cm}
%%%%%%%%%%%%%%%%%%%%%%%%%%%%%%%%%
\subsection*{\d Light hybrids}
\vspace*{-0.cm}
%%%%%%%%%%%%%%%%%%%%%%%%%%%%%%%%5
They are described by the mixed light quark- gluon  operator:
\vspace*{-0.15cm}
\beq
{\cal O}_\mu^{V(A)}=g\bar \psi \gamma_\mu(\gamma_\mu\gamma_5) \lambda_a G^a\psi, 
\eeq
\vspace*{-0.15cm}
where the hadron coupling is normalized as: 
$
\la 0\vert{\cal O}_\mu^{V(A)}\vert H\ra=\epsilon_\mu f_H M_H^2. 
$
In the pioneer works\,\cite{BALITSKY}, some controversies on the QCD expression of the two-point functions have been resolved in Ref.\,\cite{LATORRE}. From this corrected expression, one has obained the predictions\,\cite{SNB2,LATORRE} :
\vspace*{-0.15cm}
\beq
\hspace*{-0.5cm} M_{\tilde\rho}(1^{-+})=(1.4\sim 1.6)\,{\rm GeV},~~M_{\tilde\eta}(0^{--})\simeq 3.8\,{\rm GeV}.
\eeq
The mass of the exotic $0^{+-}$ hybrid related to the current:
\vspace*{-0.3cm}
\beq
{\cal O}_\mu(V)=g\bar \psi\gamma_\mu\gamma_5 \lambda_a \tilde G^a\psi,
\eeq
%\vspace*{-0.15cm}
 has been also found for Gaussian sum rule within a two-resonance model to be\,\cite{STEELES}:
 \vspace*{-0.15cm}
\beq
M_1(0^{+-}) \simeq 2.6\,{\rm GeV},~~M_2(0^{+-})\simeq  3.57 \,{\rm GeV}. 
\eeq
\vspace*{-0.15cm}
%%%%%%%%%%%%%%%%%%%%%%%%%%%%%%%%%%%%%%%
\begin{figure}[hbt]
%\vspace*{-0.5cm}
\begin{center}
\includegraphics[width=5cm]{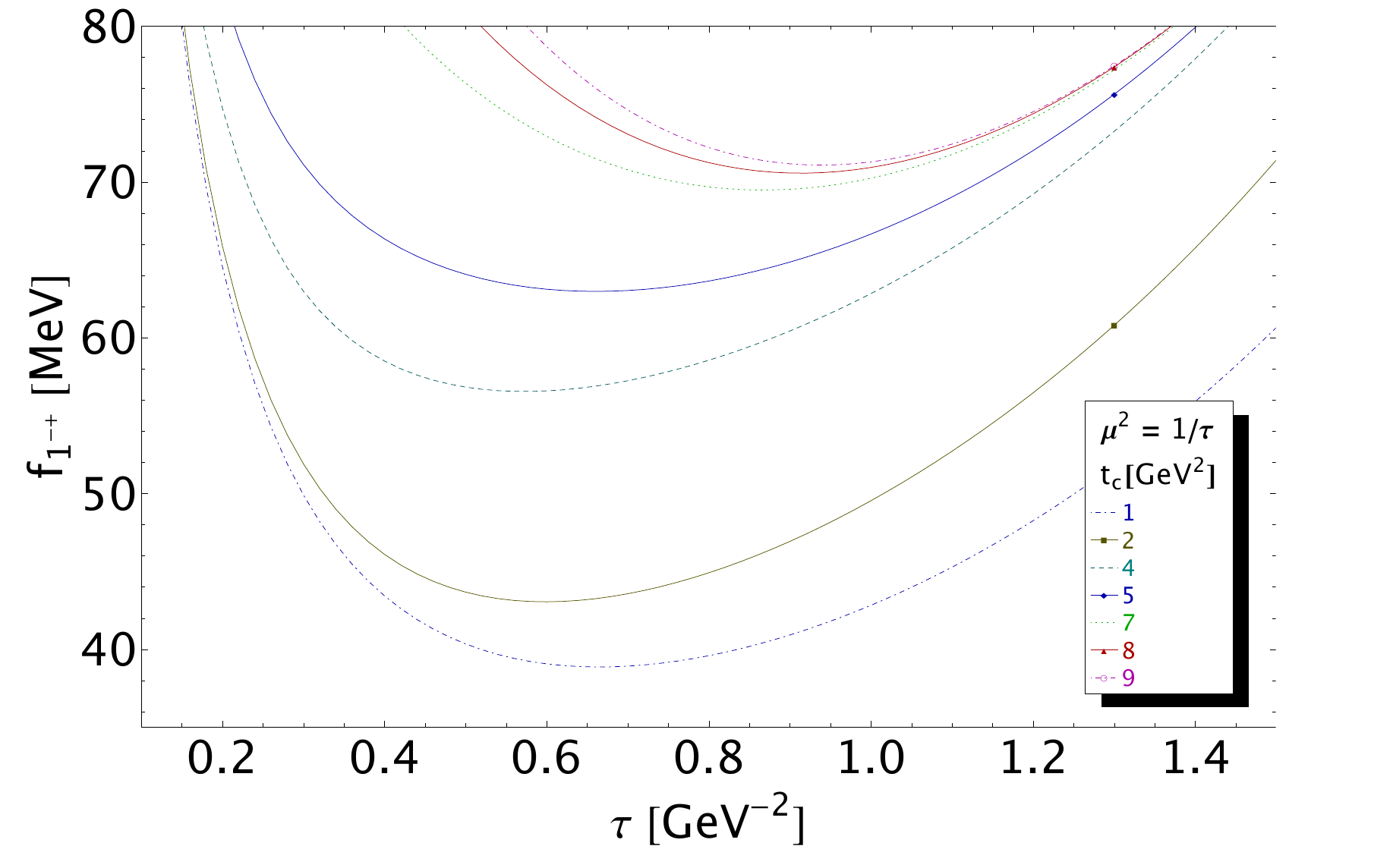}
\vspace*{-0.25cm}
\caption{\footnotesize  Behaviour of the $1^{-+}$ light hybrid coupling from the moment ${\cal L}_1$ versus $\tau$ for different values of $t_c$. .} 
\label{fig:1-coupling}
\end{center}
%\vspace*{-0.75cm}
\end{figure} 
%%%%%%%%%%%%%%%%%%%%%%%%%%%%%%%%%%%%%%%%% 
\vspace*{-0.5cm}
%%%%%%%%%%%%%%%%%%%%%%%%%%%%%%%%%%%%%%%
\begin{figure}[hbt]
\vspace*{-0.25cm}
\begin{center}
\includegraphics[width=6cm]{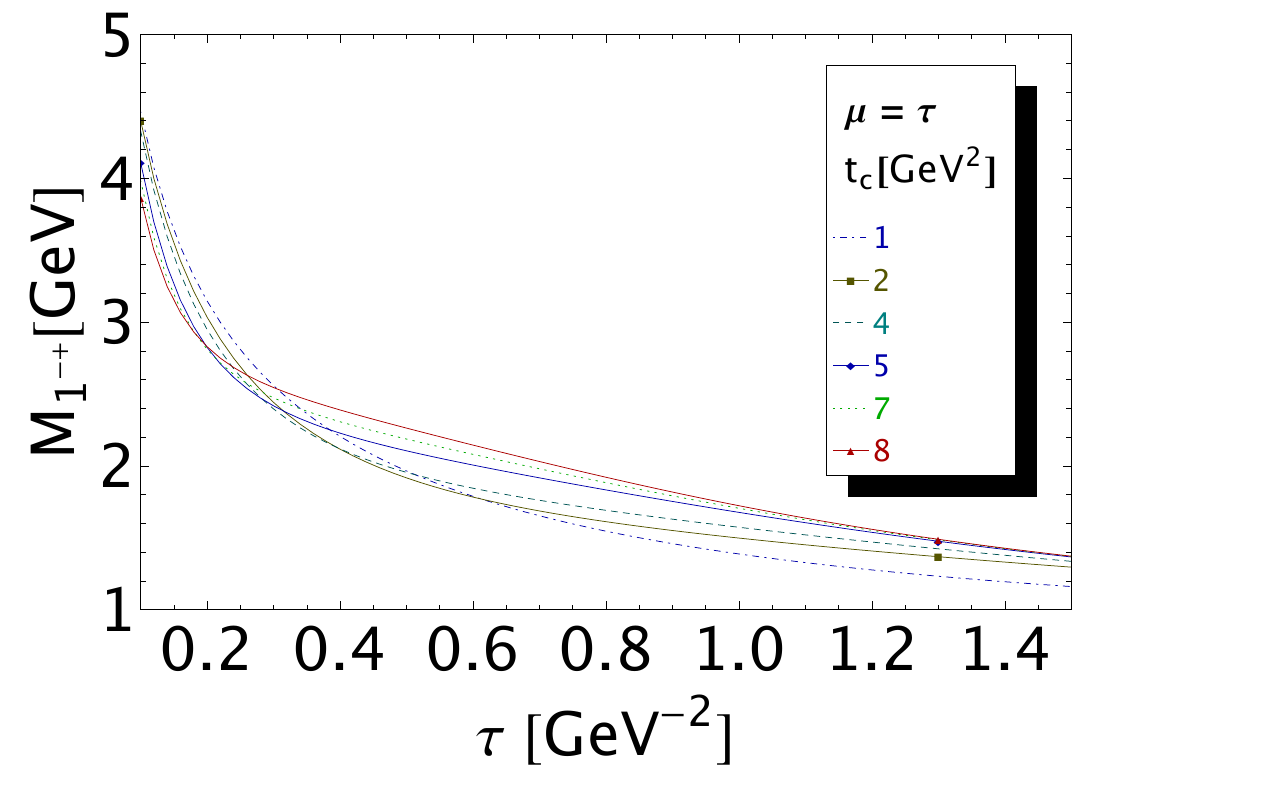}
\vspace*{-0.25cm}
\caption{\footnotesize  Behaviour of the $1^{-+}$ light hybrid mass from the moment ${\cal R}_{21}$ versus $\tau$ for different values of $t_c$. .} 
\label{fig:1-mass}
\end{center}
\vspace*{-0.25cm}
\end{figure} 
%%%%%%%%%%%%%%%%%%%%%%%%%%%%%%%%%%%%%%%%% 
%%%%%%%%%%%%%%%%%%%%%%%%%%%%%%%%%%%%%%%
\begin{figure}[hbt]
%\vspace*{-0.5cm}
\begin{center}
\includegraphics[width=5cm]{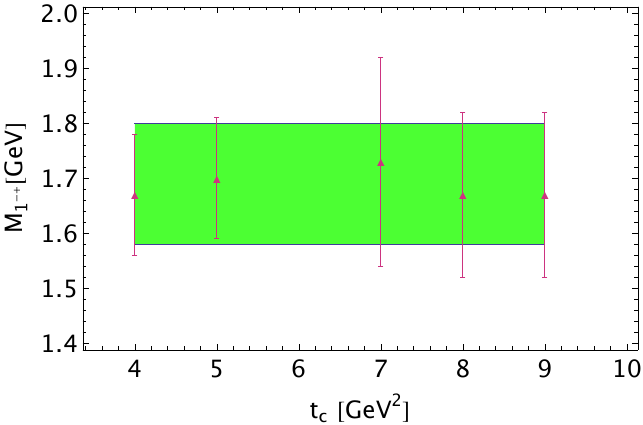}
\vspace*{-0.25cm}
\caption{\footnotesize   $1^{-+}$ light hybrid mass from the moment ${\cal R}_{21}$ at the optimal values of  $\tau$ versus $t_c$ .} 
\label{fig:plot-mass}
\end{center}
\vspace*{-0.5cm}
\end{figure} 
%%%%%%%%%%%%%%%%%%%%%%%%%%%%%%%%%%%%%%%%% 
%%%%%%%%%%%%%%%%%%%%%%%%%%%%%%%%%%%%%%%
\begin{figure}[hbt]
%\vspace*{-0.5cm}
\begin{center}
\includegraphics[width=6cm]{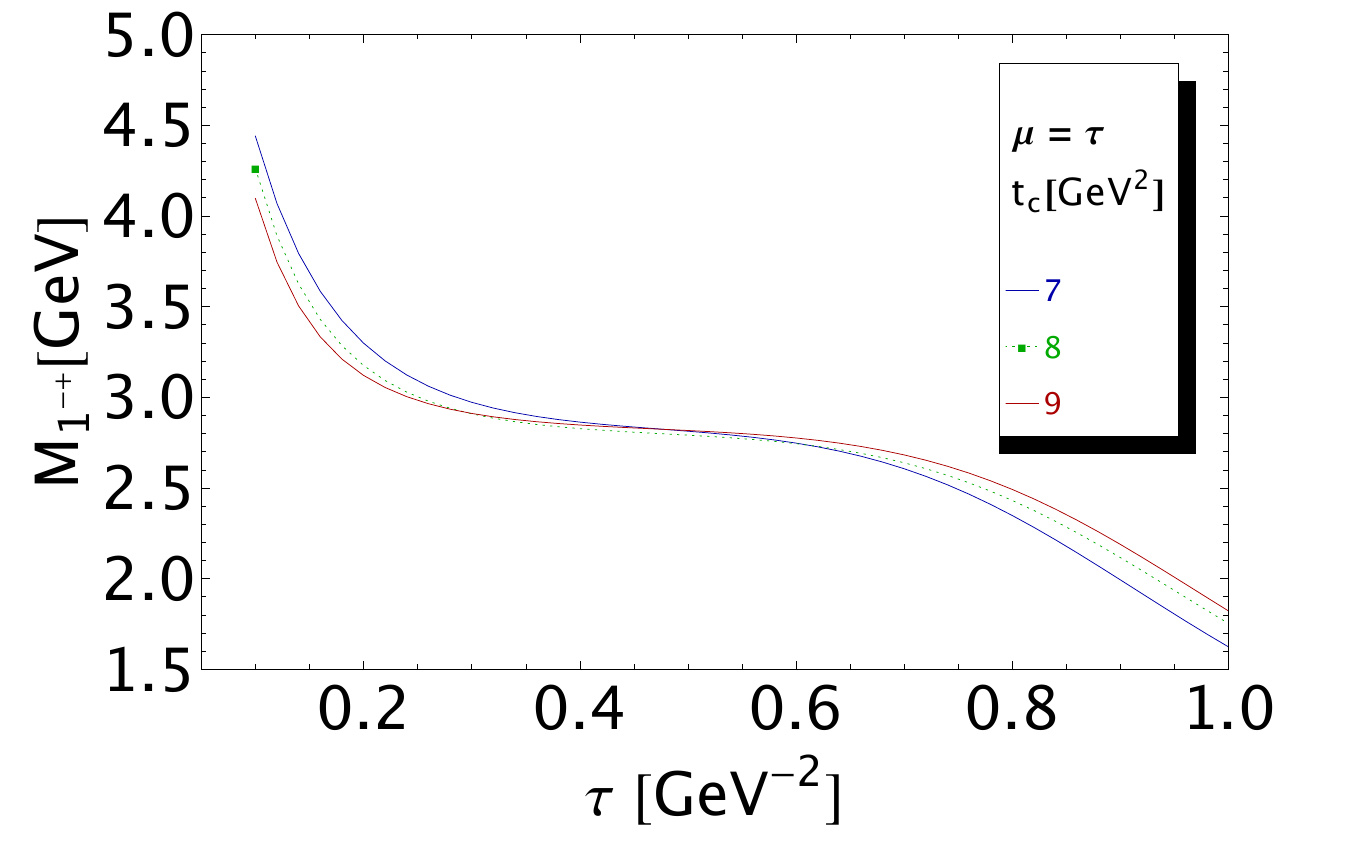}
\vspace*{-0.25cm}
\caption{\footnotesize   Mass of the $1^{-+}$ light hybrid 1st radial excitation from the moment ${\cal R}_{21}$ versus  $\tau$for different values of $t_c$. } 
\label{fig:hybrid-rad}
\end{center}
\vspace*{-0.5cm}
\end{figure} 
%%%%%%%%%%%%%%%%%%%%%%%%%%%%%%%%%%%%%%%%% 

The $1^{-+}$ meson parameters have been updated in Ref.\,\cite{SNCHET,SNHYBRID} by including the tachyonic gluon mass contribution where a mass about (1.6-1.8) GeV has been found. However, in Ref.\,\cite{STEELEV}, it is claimed that the use of the new value of the $\la g^3G^3\ra$ gluon condensate from charmonium sum rules (see Table\,\ref{tab:param}) shifts the mass to (1.72-2.60) GeV. However, as previously mentioned in Refs.\,\cite{SNB2,SNCHET}, the authors use a sum rule which is sensitive to the value of the two-point function subtraction\,\cite{BALITSKY}:
\vspace*{-0.15cm}
\beq
q^2\Pi^{(1)}_V(0)\vert_{q^2=0}\approx \frac{16\pi}{9}\alpha_s\la\bar \psi\psi\ra^2,
\eeq
\vspace*{-0.15cm}
 which cancels the contribution of the four-quark condensate in the OPE.  This effect is not taken into account  in the analysis. To avoid such ambiguities , we work with higher moments where the contribution of the $D=6$ condensates disappear to LO in the sum rule. This can be achieved by working with $q^{2n}\Pi(q^2)$ for $n\geq 1$. The analysis is shown in Fig\,\ref{fig:1-coupling}  for the coupling from the moment ${\cal L}_{1}$ and in Fig\,\ref{fig:1-mass} for the mass from the ratio ${\cal R}_{21}$. The coupling exhibits nice $\tau$-stability (the same from  ${\cal L}_2$)  but not ${\cal R}_{21}$. We extract the optimal value of the mass at the minimum of ${\cal L}_{1,2}$. The mean from the two determinations of the mass using $\tau$ from ${\cal L}_{1,2}$ are shown in Fig.\,\ref{fig:plot-mass} for $t_c\geq M_{1^{-+}}$. A similar analysis is done for the coupling. We deduce the most reliable determinations independent on the $d=6$ condensates:
\beq
\hspace*{-0.5cm}M_{\tilde\rho}=(1.69\pm 0.13)~{\rm GeV},~f_{\tilde \rho}\simeq  (52\pm 22)~{\rm MeV},
\eeq
for ${t_c}=(4\sim 9)$ GeV$^2$, where the stability is reached. 
It confirms and improves  previous predictions\,\cite{LATORRE,SNB2,SNCHET,SNHYBRID,STEELEV}. The errors have been added quadratically. For $\Delta M_{\tilde\rho}$ (resp. $\Delta f_{\tilde\rho}$),  0.11 (resp. 0.2 MeV)  is due to the sum rule analysis while 0.07 GeV (resp. 22 MeV) comes from QCD. We have used $\Lambda=(340\pm 28)$ MeV. We note that the inclusion of the tachyonic gluon mass\,\cite{SNCHET,SNHYBRID} dereases the mass by 20 MeV and increases the  coupling by 10 MeV. 
Using a two resonance model and subtracting the lowest ground state contribution, we obtain from ${\cal R}_{21}$ a nice inflexion point for $\tau\simeq (0.3\sim 0.6)$ GeV$^{-2}$(see Fig.\,\ref{fig:hybrid-rad}) and a minimum for the coupling from ${\cal L}_2$ similar to 
Fig.\,\ref{fig:1-coupling} at $\tau\simeq (0.24\sim 0.28)$ GeV$^{-2}$. This leads to the result for the 1st radial excitation:
\vspace*{-0.1cm}
\beq
\hspace*{-0.5cm}M_{\tilde\rho'}\simeq (2.8\pm 0.5)~{\rm GeV},~~~~f_{\tilde \rho'}\simeq  (37\pm 11)~{\rm MeV}.
\eeq
\vspace*{-0.1cm}
The previous results confirm  that the $\pi_1(1600)$ is a good $1^{-+}$ hybrid state candidate\,\cite{SNHYBRID} but not the $\pi_1(2050)$.
% is According to the vertex sum rule analysis, the $1^{-+}$ hybrid state is expected to decay copiously into $\rho\pi$ in contrast to the light cone sum rule result\,\cite{STEELEV}. 
%%%%%%%%%%%%%%%%%%%%%%%%%%%%%%%%%
\vspace*{-0.25cm}
\subsection*{\d Heavy-Light hybrids}
\vspace*{-0.cm}
%%%%%%%%%%%%%%%%%%%%%%%%%%%%%%%%5
The spectra of heavy-light hybrids have been studied in Refs\,\cite{GOVAERTS} and revised in\,\cite{SNB2,STEELEHL}. The unmixed states are expected to be in the range 3.40 to 5.07  (resp. 7.01 to 8.6) GeV for charm (resp. bottom) channels where there are discrepancies to be clarified for some results of  the two groups. Mixing with ordinary mesons is expected to shift upward the mass by about 0.02 to  0.05 (resp. 0.19 to 0.74) GeV for the charm (resp. bottom) channels\,\cite{STEELEHL}. 
%%%%%%%%%%%%%%%%%%%%%%%%%%%%%%%%%
\vspace*{-0.25cm}
\subsection*{\d Heavy hybrids}
\vspace*{-0.cm}
%%%%%%%%%%%%%%%%%%%%%%%%%%%%%%%%5
The spectra of heavy hybrids have been studied in Refs\,\cite{GOVAERTSH} and revised 
in Ref.\,\cite{SNB2,STEELEH}. The lightest (resp. heaviest)  charmonium state is the $1^{--}(3.36)$ (resp. $0^{--}(5.51))$.  The analogue for the bottomium states are the $1^{--}(9.70)$ (resp. $0^{--}(11.48)$). In this analysis, the negative parity states $J^{PC}=(0,1,2)^{-+}$ and $1^{--}$ are lighter than states with positive $C$ and $P$ parities $J^{PC}=(0,1)^{-+} , (0,1,2)^{++}$\,\cite{STEELEH}. 
%%%%%%%%%%%%%%%%%%%%%%%%%%%%%%%%%
\vspace*{-0.3cm}
\section{Heavy Four-quark / Molecule  sum rules}
\vspace*{-0.25cm}
%%%%%%%%%%%%%%%%%%%%%%%%%%%%%%%%%

\d Spectra of compact four-quarks and molecules have been also extracted from QSSR  two-point functions while there are some attempts to determine their widths using vertex or light cone sum rules\,\cite{NIELSEN,CHEN2}. 

\d Though phenomenologically successful for predicting the $X,T,Y,Z,DK$, fully charmed,... spectra compared with the data, the common weakness of the existing works, besides the (handwaving) choice of the sum rule window,  is the use of 
% which reminds the early days of the SVZ sum rules :
perturbative LO QCD expression where the definition of the heavy quark mass which plays a crucial role is ill-defined. The common prefered (but unjustified) choice of different authors is the $\overline{MS}$ running mass instead of the pole/on-shell quark mass which would (a priori) be more natural here as the spectral function is  evaluated within an on-shell %renormalization 
scheme. 
 % which is not justified at all without adding radiative corrections which are not easy to calculate. 
 
\d  An exception is the series of works in Refs.\,\cite{MOLE16,Zc} where the factorised NLO PT contributions have been considered and where well-founded stability criteria have been used for extracting the optimal results. In this analysis, it has been found that the NLO corrections are (in most cases) small for the mass due to some partial cancellations in the ratio of moments. This  explains (a posteriori) the success of the LO analysis for the mass predictions. However, the NLO corrections can be large for the decay constants and affect the width predictions.  

\d One can notice tha the method cannot distinguish accurately the compact four-quark states from the molecule ones leading Refs.\,\cite{MOLE16,Zc} to identify the experimental  states with the {\it Tetramole} states having their mean masses.
% are the mean of the  ones of the compact quark and molecule states. 
%%%%%%%%%%%%%%%%%%%%%%%%%%%%%%%%%
 \vspace*{-0.5cm}
\section{Concluding comments}
\vspace*{-0.22cm}
%%%%%%%%%%%%%%%%%%%%%%%%%%%%%%%%%

\d Within the last ten years, there has been intensive horizontal applications of QSSR  for extracting the hadron masses and couplings from two-point functions built with local  currents and their widths using vertex or light cone sum rules. Unfortunately, only some few new vertical applications for a much deeper improvements of the sum rules have been done.

\d One can also notice the inflation of including higher dimension condensates (sometimes until D=12) which may be a good point if all contributions of a given dimension are included and if a care has been properly done on the mixing of these given dimension condensates under renormalization as illustrated by the case of $D=6$ condensates\,\cite{SNTARRACH}.  One may also expect a large violation of the vacuum saturation estimate of these condensates like in the case of four-quark ones. Instead, it is more useful to include NLO corrections to the PT and $D\leq 6$ condensates contributions and to check carefully some Wilson coefficients not automatically generated by the quark propagator in external gluon fields. 

\d Another weak point is the optimization procedure based on the sum rule window of SVZ. Manyt results are alos extracted at the lowest value of $t_c$ where the result still increases with $t_c$. Besides this point, the error analysis is often done in a sloppy way where more details  sources are not provided.  

\d Some authors continue also to use old obsolete estimates of the QCD parameters obtained by SVZ which (to my opinion) SVZ themselves will not consider seriously at present . One should be aware that, since 1979,  a lot of efforts have been devoted to improve their values, while during the 1980-90 period, several works have improved our understanding of the sum rules. However, reading recent papers, one has the impression that no theoretical progress  has been done since the SVZ discovery and the clock has stopped in 1979 ! 

\d To my personal opinion, QCD spectral sum rules can still have a long lifetime for studying successfully the properties of hadrons and for extracting the QCD parameters, provided, that we continuously improve the method by doing a more careful job ! 

\d QSSR predictions are based on QCD first principles and complement the Lattice calculations. They are often successful and have been  obtained many years before the lattice ones !
\vspace*{-0.4cm}
%\newpage
%%%%%%%%%%%
%%%%%%%%%%%%%%%

%%%%%%%%%%%
\end{document}